\documentclass[aps,pre,twocolumn,superscriptaddress,amsmath,amssymb]{revtex4-1}
\usepackage{bm}
\usepackage{amssymb}
\usepackage{colordvi}
\usepackage{graphicx}
\usepackage{color}
\usepackage{hyperref}

\newcommand{\ov}[1]{\overline{{#1}}}
\newcommand{\be}{\begin{equation}}
\newcommand{\ee}{\end{equation}}
\newcommand{\bea}{\begin{eqnarray}}
\newcommand{\eea}{\end{eqnarray}}

\newcommand{\vep}{\varepsilon}

\newcommand{\ome}{\omega}
\newcommand{\Ome}{\Omega}

\def\Re {\mbox{Re}}

\def\nn{\nonumber}

\begin{document}

\title{Photonic Crystal Architecture for Room Temperature Equilibrium
  Bose-Einstein Condensation of Exciton-Polaritons}

\date{\today}

\author{Jian-Hua Jiang}
\affiliation{Department of Physics, University of Toronto, Toronto,
  Ontario, M5S 1A7 Canada}
\author{Sajeev John}
\affiliation{Department of Physics, University of Toronto, Toronto,
  Ontario, M5S 1A7 Canada}
\affiliation{Department of Physics, King Abdulaziz University, Jeddah,
  Saudi Arabia}

\begin{abstract}
We describe photonic crystal microcavities with very strong
light-matter interaction to realize room-temperature, equilibrium,
exciton-polariton Bose-Einstein condensation (BEC). 
This is achieved through a careful balance between strong
light-trapping in a photonic band gap (PBG) and large exciton density
enabled by a multiple quantum-well (QW) structure with moderate
dielectric constant. This enables the formation of long-lived, dense
10~$\mu$m - 1~cm scale cloud of exciton-polaritons with vacuum Rabi
splitting (VRS) that is roughly 7\% of the bare exciton recombination
energy. We introduce a woodpile photonic crystal made of
Cd$_{0.6}$Mg$_{0.4}$Te with a 3D PBG of 9.2\% (gap to central frequency
ratio) that strongly focuses a planar guided optical field on CdTe 
QWs in the cavity. For 3~nm QWs with 5~nm barrier
width the exciton-photon coupling can be as large as
$\hbar\Ome=$55~meV (i.e., vacuum Rabi splitting
$2\hbar\Ome=$110~meV). The exciton recombination energy of 1.65~eV
corresponds to an optical wavelength of 750~nm. For $N=$106 QWs 
embedded in the cavity the collective exciton-photon coupling per QW,
$\hbar\Ome/\sqrt{N}=5.4$~meV, is much larger than state-of-the-art
value of 3.3~meV, for CdTe Fabry-P\'erot microcavity. The maximum BEC
temperature is limited by the depth of the dispersion minimum for the
lower polariton branch, over which the polariton has a small effective
mass $\sim 10^{-5}m_0$ where $m_0$ is the electron mass in
vacuum. By detuning the bare exciton recombination energy above the 
planar guided optical mode, a larger dispersion depth is achieved,
enabling room-temperature BEC. The BEC transition temperature ranges
as high as 500~K when the polariton density per QW is increased to   
$(11a_B)^{-2}$ where $a_B\simeq 3.5$~nm is the exciton Bohr radius and the
exciton-cavity detuning is increased to 30~meV. A high quality PBG can
suppress exciton radiative decay and enhance the polariton lifetime to
beyond 150~ps at room temperature, sufficient for thermal equilibrium
BEC. 

\end{abstract}



\maketitle

\section{Introduction}
Semiconductor microcavities offer an unique platform to study
fundamental phenomena of quantum electrodynamics in the strong
coupling regime\cite{weisbuch1,koch,book}. Thanks to advances in
growth and manipulation technologies, both excitons and
cavity photons are tunable in a variety of ways which enable
observations of novel phenomena\cite{book,rmp,rmp2}. In 
the last decade non-equilibrium exciton-polariton BEC have been
observed in 1D Fabry-P\'erot microcavities in various 
materials\cite{deng-science,bec1,bec3,bec4,rmp,rmp2}. However,
polariton lifetime is often too short to achieve thermal equilibrium 
BEC. Moreover, the BEC temperature is limited by the weakness of
photon confinement and the resulting insufficient exciton-photon
coupling strength.

In an optical cavity, an atomic excitation and a photon may couple
together to form a coherent superposition when their decay
rates (times $\hbar$) are much smaller than their coupling.
In a semiconductor microcavity\cite{weisbuch1,koch,book}, excitons and photons
constitute polaritons in this manner. The quantum
superposition of excitons and photons confers them with remarkable
dynamics. For instance, the polariton effective mass, inherited from
the cavity photon, is about $10^{-10}$ times the mass of Rubidium atom.
Consequently, the BEC transition temperature of polariton gases is
usually above 1~K. In the last decade signatures of 
non-equilibrium polariton BEC has been reported in semiconductor
microcavities\cite{deng-science,bec1,bec3,bec4,rmp,rmp2}. Although
aspects of macroscopic coherence and superfluidity have
been observed/claimed\cite{deng-science,superfluid,vort1,joseph,soliton},
the polariton lifetime has been too short (about
1~ps) to achieve genuine thermal equilibrium\cite{butov}.
Recently high-quality-factor Fabry-P\'erot microcavities have
provided longer polariton lifetime\cite{snoke3}.
Nevertheless, polaritons can radiatively decay into extraneous and
degenerate optical modes of these 1D periodic structures from a
micron-scale confinement region.

Due to the half-photon and half-exciton nature of the polariton, the
Bose condensate contains polariton-polariton interactions (i.e.,
optical non-linearity). Such a nonlinear, coherent light source has
potential application as a laser with low excitation threshold and
novel photon statistics\cite{atac,deng-science,bec1,electric,yang3}.
Nonlinearity can also be exploited for the development of 
all-optical transistors\cite{transistor},
diodes\cite{diode}, and
switches\cite{switch-th,fast-switch}. Unfortunately, most realizations
of polariton BEC have been at cryogenic temperatures. Above
room-temperature polariton BEC\cite{rt} is essential for broad
practical applications. While studies on 
GaN\cite{gan}, ZnO\cite{zno,zno2}, and organic materials\cite{organic}
have found room-temperature polariton light emission and lasing, these
are not sufficiently long-lived for equilibrium BEC. These room
temperature effects are facilitated by large exciton binding energy
$E_b$ and exciton-photon coupling $\hbar\Ome$. With small exciton Bohr
radius $a_B$, the saturation density $\simeq (5a_B)^{-2}$ at which
electron-hole pairs unbind due to many-body effects (screening and
phase-space filling as revealed in Ref.~\cite{saturation}) is likewise
very high in these materials. On the other hand, small excitons are
very sensitive to local disorder. This leads to strong inhomogeneous
broadening that smears the exciton and polariton dispersion. The
inhomogeneous broadening in the polymer material measured from optical
spectroscopy in Ref.~\cite{organic} is as high as 60~meV while the
exciton-photon coupling is $\hbar\Ome=$58~meV. The inhomogeneous
broadening in GaN and ZnO are around 16~meV\cite{gan-ib} and
25~meV\cite{zno-ib}, respectively, while the exciton-photon coupling
are 25~meV\cite{gan-coupl} and 29~meV\cite{zno-coupl}, respectively. In
contrast the inhomogeneous broadening in GaAs and CdTe is around
1~meV\cite{exp-dEx}. Nevertheless the exciton-polariton lifetime is
still below 1~ps\cite{zno2} in these materials due to the nature of
the optical cavity utilized. {Recently, macroscopic occupation of
  photonic states in microcavities filled with dye molecules has been
  observed at room temperature\cite{weitz1,weitz2}. This
  quasi-condensate\cite{snoke-review} in the weak-coupling regime
  exhibits strongly fluctuating photon number\cite{weitz2}, unlike BEC
  of polaritons in the strong-coupling regime with conserved particle
  number\cite{rmp}.}




\begin{figure}[]
  \centerline{\includegraphics[height=3.8cm]{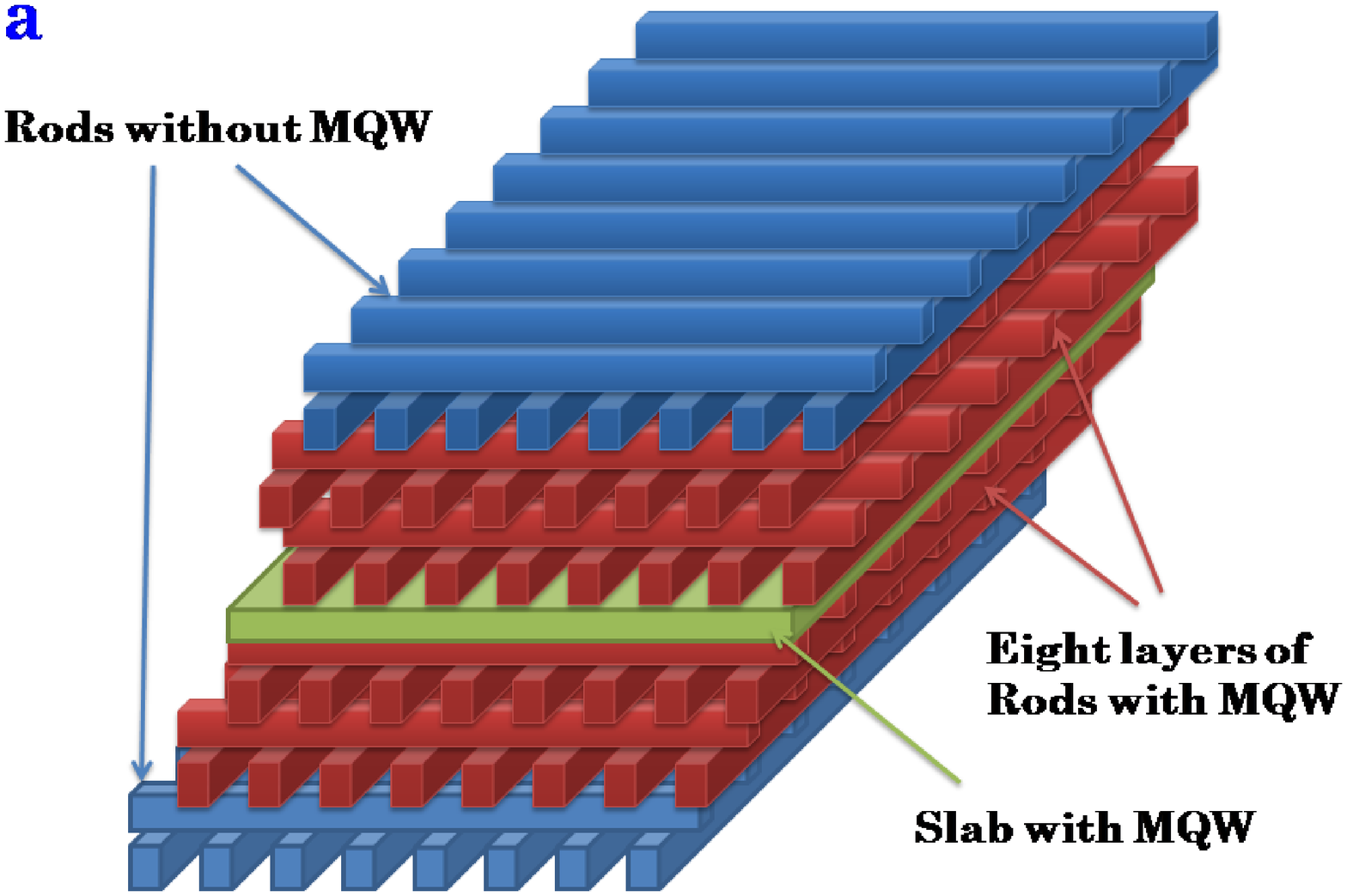}}
  \centerline{\includegraphics[height=4.1cm]{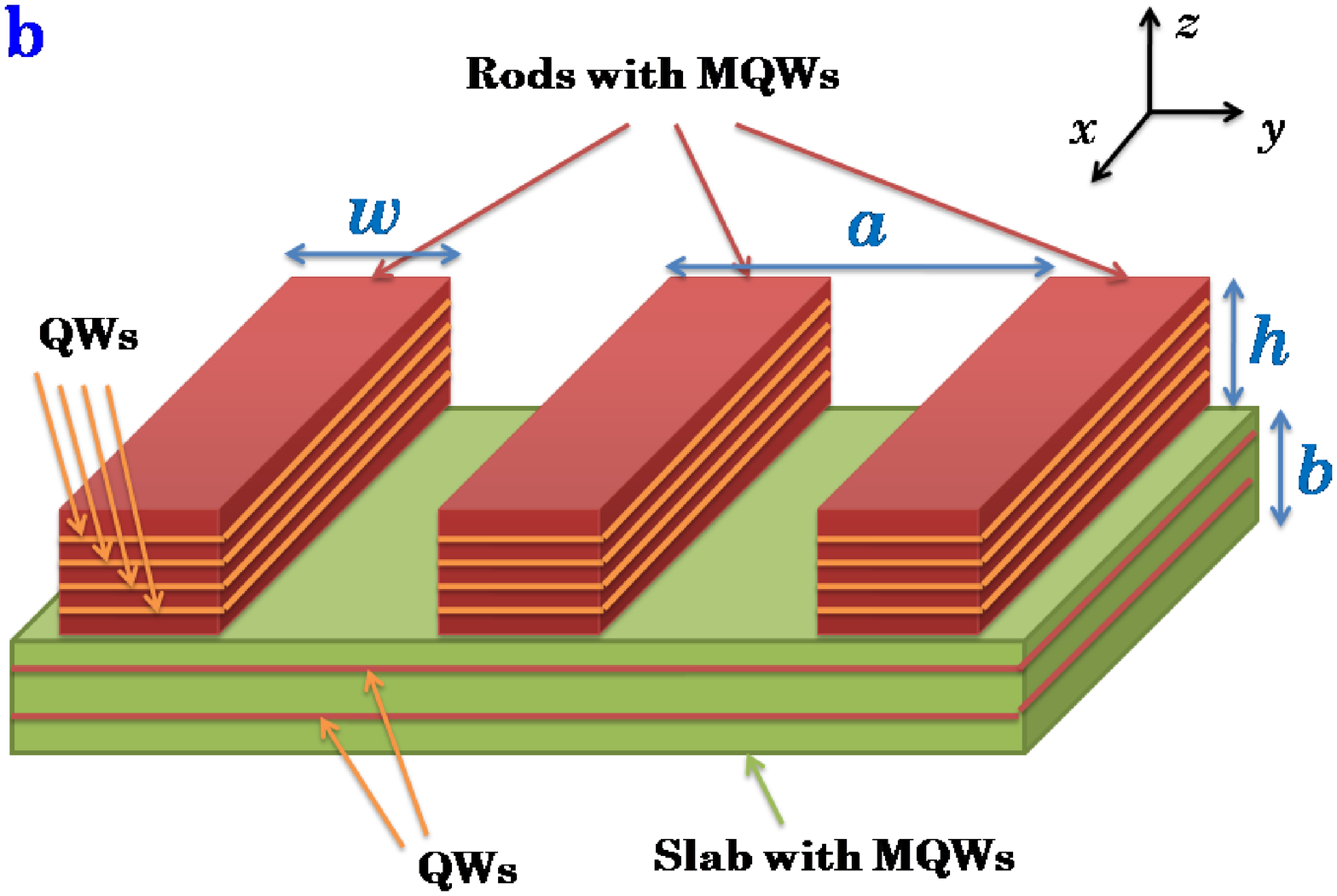}}
  \centerline{\includegraphics[height=3.8cm]{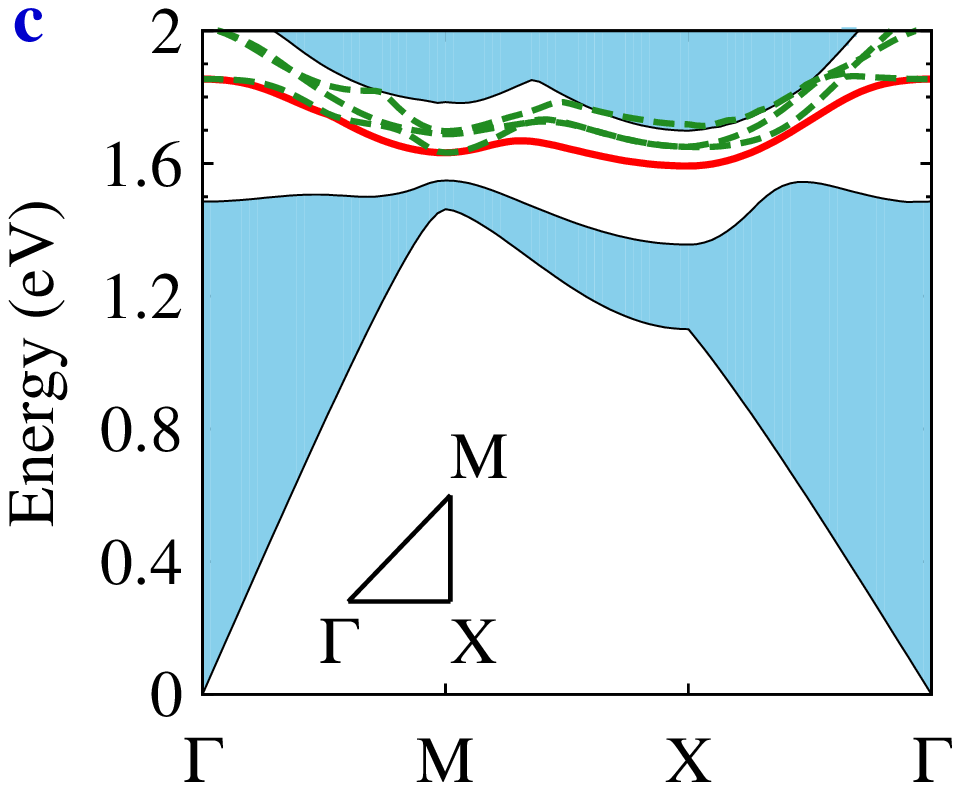}\includegraphics[height=3.8cm]{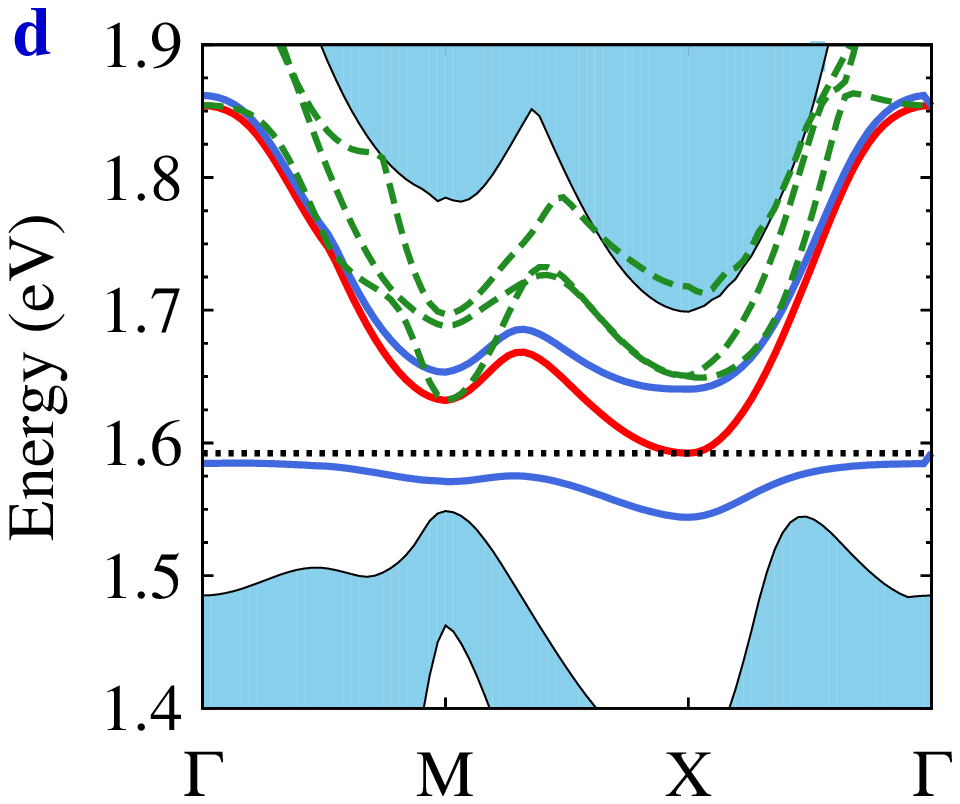}}
  \caption{ Photonic crystal architecture and
      photonic band dispersion. (a) A slab layer sandwiched between
    two woodpile photonic crystals induces 2D photonic bands (``guided modes'') in
    the 3D PBG with $\delta \ome/\ome_c = 9.2\%$
    where $\delta \ome$ is band gap width and $\ome_c$ is
    the PBG central frequency. (b) QWs are grown in the slab and
    the eight nearby layers of rods. Typically $a\simeq 360$~nm,
    $w=0.25a\simeq 90$~nm, $h=0.3a\simeq 110$~nm, $b=0.06a\simeq 20$~nm. There are about
    100 QWs with the same width. There are two QWs in the slab layer and about
    12 QWs in each rod. The QW width is chosen to be 3, 4, 5,
    6~nm for comparison, while the width of the barrier layers between
    QWs is fixed to be 5~nm. QWs are made of CdTe (dielectric constant
    8.5) while the barriers and photonic crystals are made of
    Cd$_{0.6}$Mg$_{0.4}$Te (dielectric constant 7.5). In the regions with QWs, the
    dielectric constant is taken to be 8 (average of
    the two materials). (c) and (d) Dispersion of the 2D guided modes 
    (green dashed and red solid curves) and the bulk 3D photonic
    bands (shaded regions). The first Brillouin zone of the 2D
    guided modes is $-\frac{\pi}{a}\le q_x,q_y<\frac{\pi}{a}$. The
    lowest 2D photonic band (the solid curves in (c) and (d)) is
    coupled with excitons (dotted curve in (d)) in QWs. The lower
    and upper polariton branches (blue solid curves in (d)) are
    depicted for exciton recombination in resonance with the lowest 2D
    guided mode. In this case, each QW width is 4~nm; exciton recombination energy is $E_{X0}=$1.59~eV; the
    photonic lattice constant is chosen to be $a=360$~nm. The
    exciton-photon coupling strength and the on-resonance lower
    polariton dispersion depth is 48~meV. Larger dispersion depth and
    higher BEC temperature is achieved by detuning the exciton
    recombination energy above the lowest 2D guided mode.}
  \label{struc}
\end{figure}

In this article, we introduce an alternative approach, using 3D
PBG\cite{pbg1,pbg2} based
microcavities\cite{cavity-laser,noda,qd-cavity} to greatly enhance the
exciton-photon coupling and eliminate radiative decay of
polaritons. We focus on CdTe based photonic crystal microcavities
since these offer an effective balance between strong light
confinement and high exciton density. The exciton binding energy in
CdTe QWs has been measured to be about
20~meV (varying with QW width, temperature,
etc)\cite{Eb1,Eb2,sat-exp} leading to an exciton Bohr radius
$a_B$ about 4~nm. An exciton-photon coupling of
$\hbar\Ome=13$~meV has been experimentally realized in CdTe based Fabry-P\'erot
microcavities with $N=$16 QWs\cite{bec3}. These data
indicate that neither the exciton binding energy nor the
exciton-photon coupling in CdTe structures is greater than room
temperature $k_BT=26$~meV. Bare excitons in CdTe QWs
are not stable at room temperature and the polariton BEC temperature
is below room temperature in Fabry-P\'erot microcavities. Indeed
non-equilibrium polariton BEC in a CdTe based Fabry-P\'erot
microcavity is only 19~K\cite{bec3}. On the other hand, the refractive
index of CdTe is sufficient for much stronger light-trapping in the
form of 3D PBG. Here we show that using a PBG architecture, the
photonic field is much more strongly focused on suitably placed
QWs. Consequently, the exciton-photon coupling is enhanced beyond
50~meV (i.e., vacuum Rabi splitting beyond 100~meV). This stabilizes
the exciton-polariton and enables equilibrium BEC above room
temperature.

For excitons in CdTe multiple QWs (MQWs) with width 3~nm and
barrier width 5~nm, the exciton recombination energy is 1.65~eV (optical wavelength
$750$~nm) and the exciton binding energy is calculated by the
effective mass approximation\cite{harrison} to be 23~meV (larger than
the 9~meV in GaAs QW of the same width\cite{our}). The exciton Bohr
radius of 3.5~nm is likewise smaller than the corresponding radius in
GaAs (8.4~nm)\cite{our}, but larger than in ZnO (1.4~nm) and GaN
(2.8~nm)\cite{Ab}. These properties of CdTe lead to stronger coupling, higher
saturation density, and stronger thermal stability of excitons
compared to GaAs. CdTe is nevertheless suitable for high quality
microcavities for which growth technology is well-developed\cite{book,Eb2}. CdTe has a
dielectric constant of 8.5\cite{Eb2,mhhp,dielectric} above the
threshold for PBG formation. We propose to use Cd$_{0.6}$Mg$_{0.4}$Te
as the barriers between QWs as well as for the background photonic
crystal. The dielectric constant of Cd$_{0.6}$Mg$_{0.4}$Te is
7.5\cite{mgte-diel}. This gives a photonic
band gap (PBG) in the woodpile architecture (see Fig.~\ref{struc}) of $\delta
\ome/\ome_c=9.2\%$ ($\delta \ome$ denotes band gap width and $\ome_c$
is the PBG central frequency). The dielectric constant of the regions with
MQWs is taken to be 8, which is the average value of the QW and barrier
materials. Using $N=$106 QWs, the exciton-photon coupling is as
large as $\hbar\Ome=55$~meV, i.e., the vacuum Rabi-splitting is
110~meV. The exciton-photon coupling per QW of
$\hbar\Ome/\sqrt{N}=5.4$~meV is considerably larger than
$13/\sqrt{16}=3.3$~meV for the Fabry-P\'erot microcavity reported in
Ref.~\cite{bec3} (i.e., exciton-photon coupling of 13~meV for 16
QWs). We also calculate the exciton-photon coupling 
when the number of QWs in each rod is reduced by increasing the
barrier width (keeping two QWs in the central slab). 
The exciton-photon coupling is 45~meV if the barrier width is
increased to 10~nm while there are 7 QWs in each rod in the active
region (total of 58 QWs). When the barrier width is increased to 20~nm with 4 QWs in each
rod in the active region (total 34 QWs), the exciton-photon coupling is 40~meV. If
there are only two QWs in the central slab and no QW in the rods, the
exciton-photon coupling remains as high as 33~meV. All these energies are greater than
room temperature $k_BT=26$~meV and (with positive detuning of the
exciton recombination energy from the lowest 2D guided mode) are shown
below to support room-temperature polariton BEC.

Fabrication of the structure proposed 
(as shown in Fig.~\ref{struc}a and 1b) is possible using wafer-fusion and laser
beam-assisted precise alignment of
rods\cite{noda1,noda2}. It has been demonstrated\cite{noda} that a
layer of rods with MQWs (grown by molecular beam epitaxy and etched)
can be placed in a woodpile photonic crystal. By repeated
wafer-fusion, laser-assisted alignment, careful etching and
polishing (monitored by thin-film interference), sandwich
structures can be fabricated with high precision. Accurate fabrication
of GaAs woodpile photonic crystals with a near-infrared, MQW,
light-emitting layer was reported by Noda {\sl et
  al.}\cite{noda}. Similar techniques can, in principle, be employed
for our proposed CdTe photonic crystal structure.

3D photonic crystal microcavities have been fabricated with much
higher quality factor than their counterpart Fabry-P\'erot
microcavities. Quality factors as high as $2\times 10^6$ have been
demonstrated experimentally for microcavities in the woodpile
architecture\cite{high-q}. In addition, 3D PBG eliminates purely
radiative recombination of excitons into unwanted modes when the lower
polariton energy is in the PBG. In this case, the polariton lifetime
is determined by exciton non-radiative recombination. In CdTe QWs
at room temperature, the exciton photoluminescence decay time is
measured to be 150~ps in Ref.~\cite{pl-decay}, suggesting that the 
non-radiative recombination time is even longer. This time scale
is sufficient for polaritons to reach thermal equilibrium since the
exciton--optical-phonon scattering in CdTe at room temperature is
quite efficient\cite{therm}. Our calculation reveals that room
temperature equilibrium polariton BEC can be achieved for a polariton
density of $5\times 10^3$~$\mu$m$^{-2}$ (exciton density in each QW of
$\sim (30a_B)^{-2}$, below the saturation density of $(5a_B)^{-2}$
where $a_B=3.5$~nm is the bare exciton Bohr radius for QW of width
3~nm). This occurs when the exciton recombination energy is detuned
30~meV above the lowest 2D guided mode and the excitons are trapped in
a box with side length of 10-$10^{4}$~$\mu$m. Realization of equilibrium polariton
BEC at and above room temperature may open the door to broad
applications of quantum mechanics beyond the microscopic scale.

\begin{figure}[]
  \centerline{\includegraphics[height=3.8cm]{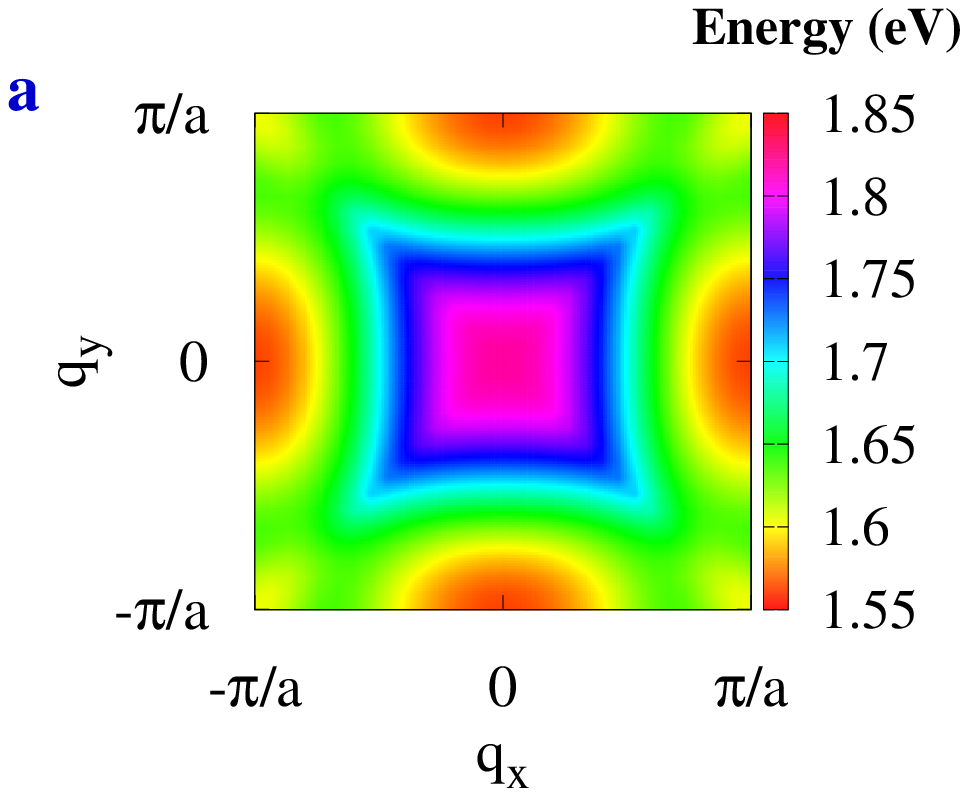}\includegraphics[height=3.8cm]{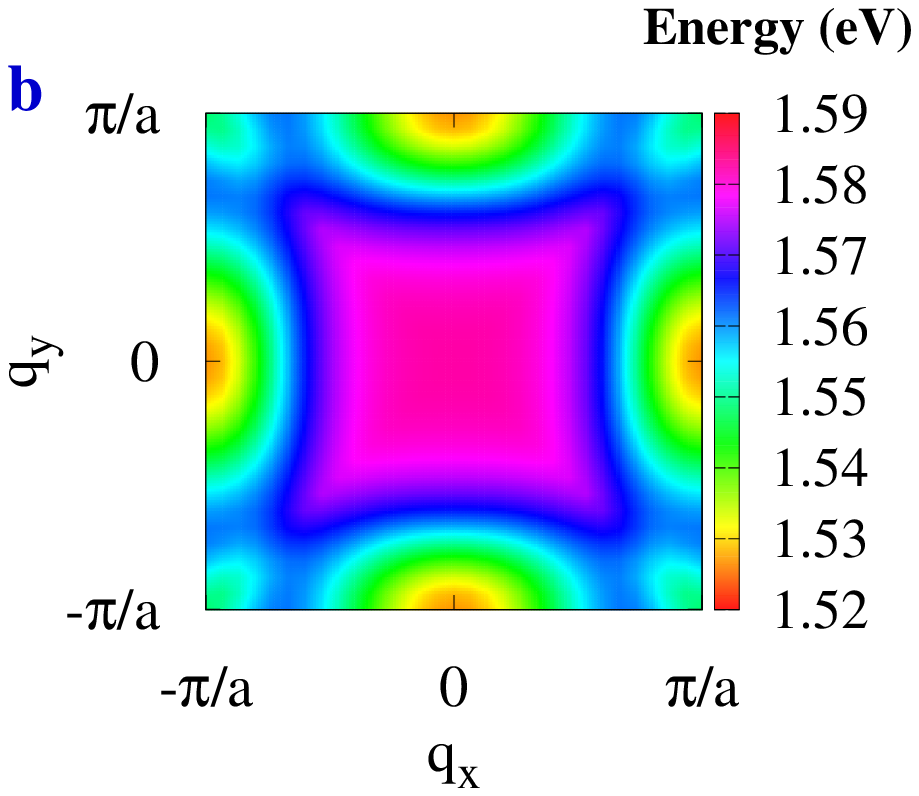}}
  \caption{ Dispersion of confined photonic and
      polaritonic bands in 2D wavevector space. (a)
    Dispersion of the lowest confined photonic band in the first
    Brillouin zone $-\frac{\pi}{a}\le q_x, q_y<\frac{\pi}{a}$. The
    lattice constant is $a=365$~nm. (b)
    Dispersion of the lower polariton branch in the same wavevector
    region. The QW width
    is 4~nm. The lower band edge (1.56~eV) of guided 2D modes within
    the 3D PBG is 30~meV below the excitonic recombination energy
    ($E_{X0} = 1.59$~eV). The exciton-photon coupling of 48~meV
    involves the collective response of 100 QWs.}
  \label{disper}
\end{figure}

\begin{figure}[]
  \centerline{\includegraphics[height=3.6cm]{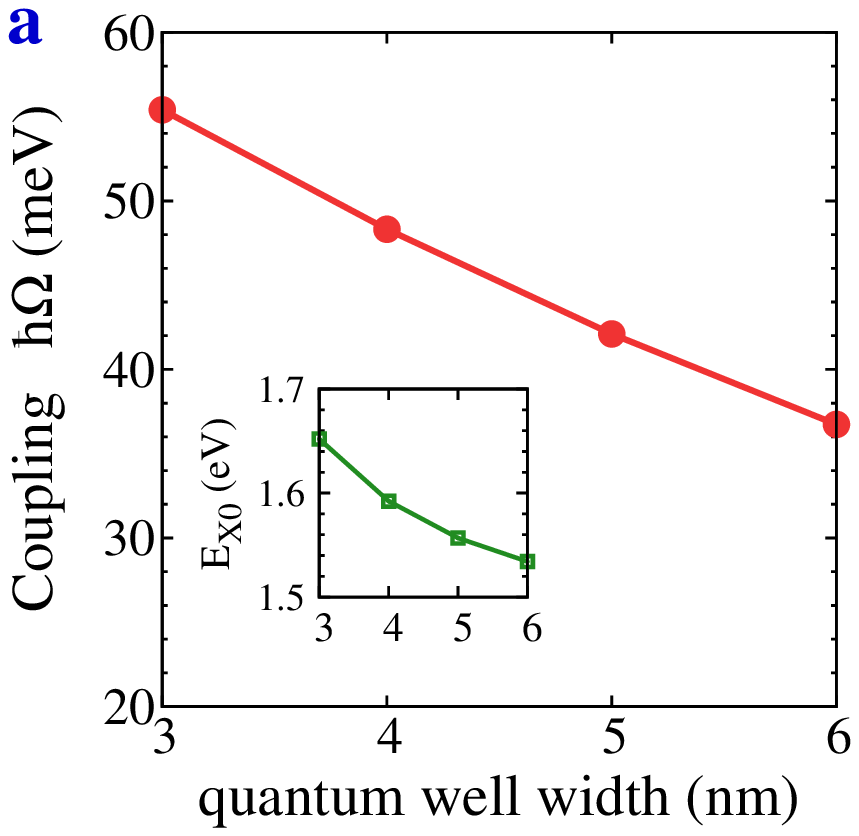}\includegraphics[height=3.6cm]{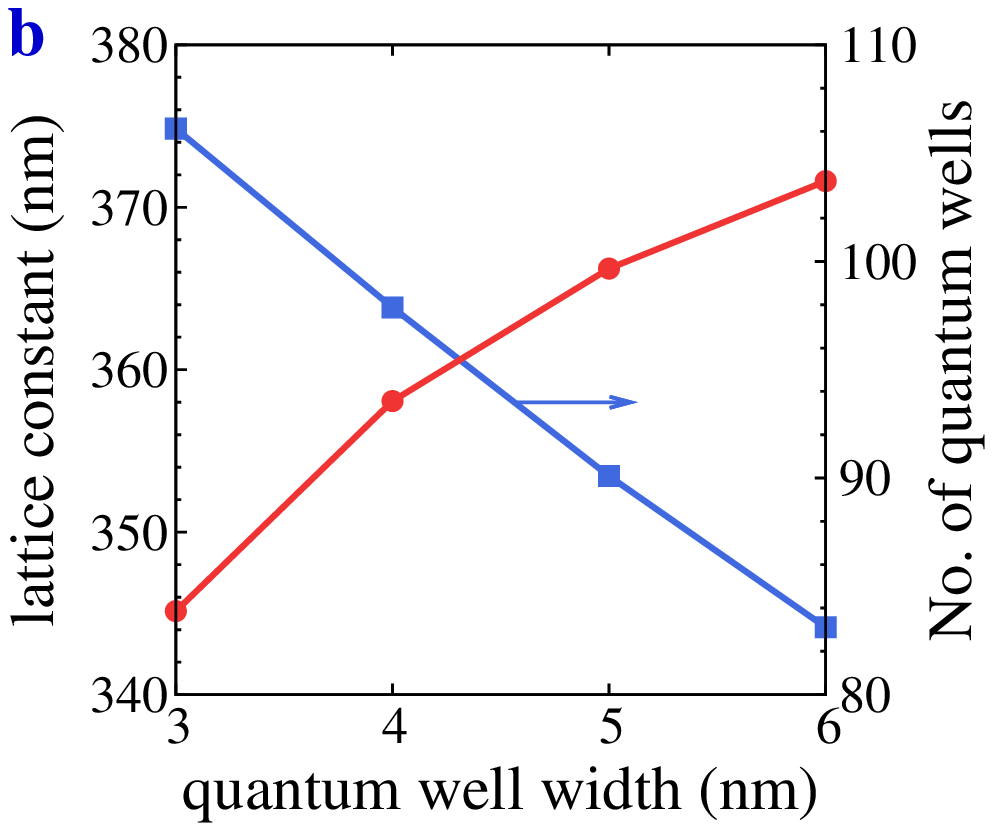}}
  \centerline{\includegraphics[height=3.6cm]{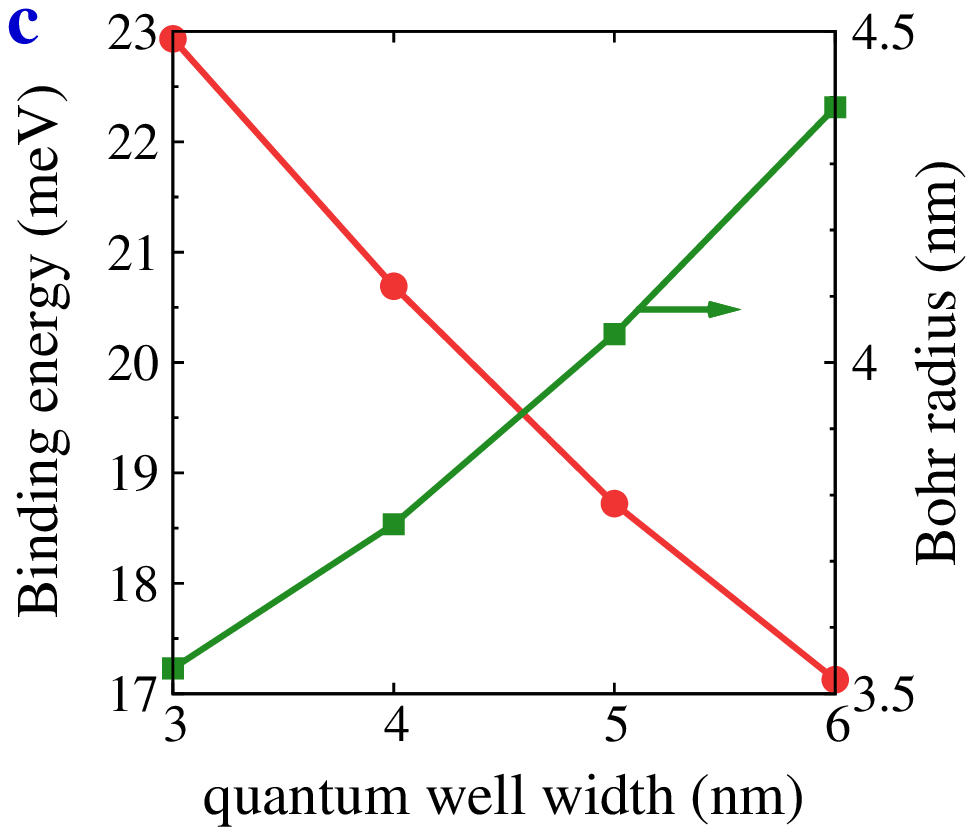}\includegraphics[height=3.6cm]{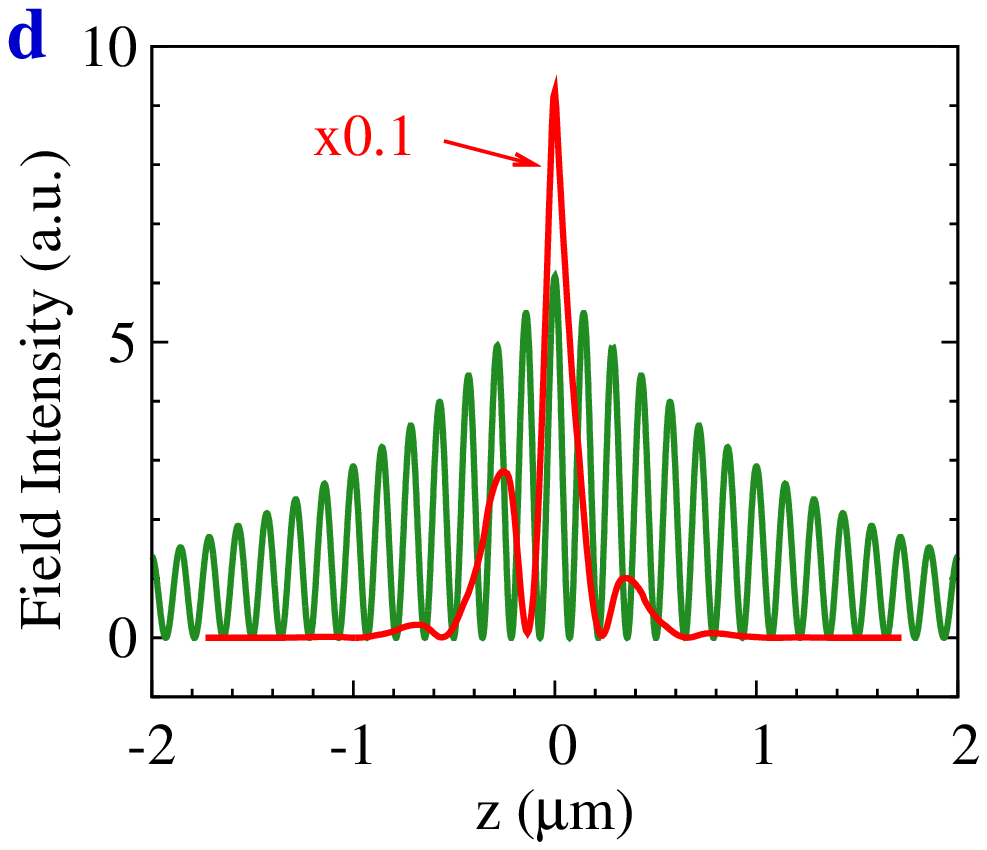}}
  \centerline{\includegraphics[height=3.cm]{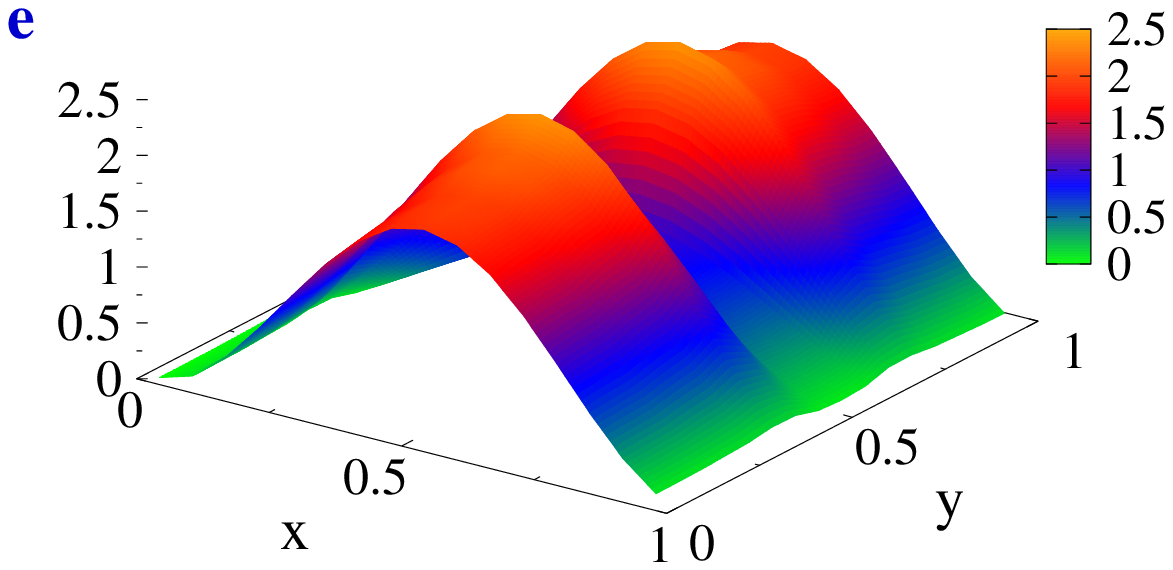}}
  \caption{ Very strong light-matter interaction in photonic band gap
    2D guided mode. (a) Exciton-photon collective interaction strength
    $\hbar\Ome$ as a function of QW width for fixed barrier width of
    5~nm. The horizontal axis spans from $N=106$ to 83 QWs. Inset: exciton
    recombination energy $E_{X0}$. The lowest 2D photonic band edge
    $\hbar\ome_0$ is kept in resonance with the exciton recombination
    energy $E_{X0}$ as the QW width is varied. The barrier layers between QWs are
    made of Cd$_{0.6}$Mg$_{0.4}$Te. 
    (b) As $E_{X0}$ varies with the QW width, the
    photonic crystal lattice constant $a$ is adjusted (red curve) to keep
    $\hbar\ome_0=E_{X0}$. Since $a$ changes, the total number $N$ of
    QWs (blue curve) in the slab and rods also changes. 
    (c) Exciton binding energy and Bohr radius for different QW
    widths calculated using effective mass
    approximation\cite{harrison} [see Appendix]. 
    (d) Average electric field intensity $S^{-1} \int d{\vec \rho} |{\vec {\cal E}}({\vec \rho},z)|^2$ profile at 2D guided mode
    edge in PBG cavity (red curve) compared to
    that in Cd$_{0.4}$Mg$_{0.6}$Te/Cd$_{0.8}$Mn$_{0.2}$Te Fabry-P\'erot cavity\cite{sat-exp}
    (green curve) along the $z$-direction. The 
    photons in both cavities have band edge energy,
    1.59~eV. The photonic field intensity in
    the PBG cavity is multiplied by a factor of 0.1 in the figure. The Fabry-P\'erot cavity is a $\lambda/2$ cavity with
    $\lambda = 780$~nm. The lattice constant of the photonic 
    crystal is $a=360$~nm. (e) Field
    intensity $|{\vec {\cal E}}|^2$ profile in the $x$-$y$ plane (in
    one unit cell of photonic crystal) at $z=0$. Here the averaged intensity is 1. $x$ and
    $y$ coordinates are in unit of the photonic lattice constant
    $a$. } 
  \label{vrs}
\end{figure}

\section{Polaritons in A 3D Photonic Band Gap microcavity}

Figs.~\ref{struc}a and \ref{struc}b depict our proposed woodpile photonic crystal
heterostructure. The distance between the nearest neighbor rods in the
same layer is denoted by $a$, the height of each rod is $h=0.3a$, the
width of each rod is $w=0.25a$, and the thickness of the solid central
slab is $b=0.06a$. In order to place the lowest 2D guided mode within
the PBG in resonance with the exciton transition at 1.59~eV (for QW
width 4~nm) typically $a=360$~nm. The PBG is centered in the range of
750~nm\cite{infrared-phc}. MQWs are grown in the slab and the eight layers of rods close to the
slab. The other regions above and below are the cladding woodpile photonic
crystals. The QWs are made of CdTe, while the barrier layers between
QWs are made of Cd$_{0.6}$Mg$_{0.4}$Te. Both materials are direct electronic
band-gap semiconductors with band structure similar to GaAs. The
electronic band-gaps of CdTe and Cd$_{0.6}$Mg$_{0.4}$Te are 1.475~eV and
2.28~eV, respectively\cite{madelung,kim} (from linear interpolation between
the band gaps of CdTe and MgTe, the latter being 3.49~eV). The
cladding rods without QWs are also made of Cd$_{0.6}$Mg$_{0.4}$Te. The structure breaks the
translation symmetry along the $z$-direction (see Fig.~\ref{struc}a), but has lattice translation symmetry in the $x$-$y$ plane (the
coordinate axes are depicted in Fig.~\ref{struc}b). There is no inversion
symmetry with respect to the slab. However the system has a $D_{2 d}$
symmetry with two mirror planes: the $y$-$z$ and $x$-$z$
planes\cite{group}. There are several 2D guided photonic bands localized around
the slab. The dispersion of those confined modes within the PBG are
shown in Figs.~\ref{struc}c and \ref{struc}d. The electric field in these 2D bands can
be written as:
\be
{\vec {\cal E}}_{i,{\vec q}}({\vec r}) = \sqrt{\frac{\hbar \ome_{i,{\vec q}}}{2\vep_0
    S}} {\vec u}_{i,{\vec q}}({\vec r}) e^{i{\vec  q}\cdot{\vec \rho}} .
\label{eeuu}
\ee
Here $\vep_0$ is the vacuum permittivity, $S$ is the area of the
structure in the $x$-$y$ plane, ${\vec r}=({\vec \rho},z)$ with ${\vec
  \rho}=(x,y)$, and $\ome_{i,{\vec q}}$ is the frequency of a photon in the
$i$-the band with a Bloch wavevector ${\vec q}$ in the 2D Brillouin
zone $-\frac{\pi}{a}\le q_x,q_y<\frac{\pi}{a}$. The field is 
normalized such that $S^{-1} \int d{\vec \rho} d z \vep({\vec
  r}) \left| {\vec u}_{i,{\vec q}}({\vec r})\right|^2 = 1$ 
where $\vep({\vec r})$ is the coordinate-dependent dielectric
function. Photons in the lowest confined photonic band and
electron--heavy-hole excitons in the MQWs interact to 
form the lower and upper polariton branches. These are depicted in
Fig.~\ref{struc}d when the exciton recombination energy coincides with 2D
photonic band edge. The dispersion
of the lowest confined photonic band and that of the lower polariton
branch in the 2D photonic Brillouin are shown in Fig.~\ref{disper} for the case
when the exciton recombination energy is detuned by 30~meV above the
2D photonic band edge.

The energy levels and wavefunctions of the $1s$ excitonic states in
QWs are obtained from diagonalizing the effective mass
Hamiltonian\cite{harrison} [see Appendix for details].
Combining the microscopic details of cavity photons and QW excitons,
the Hamiltonian of the coupled exciton-photon system\cite{our} is
given by:
\begin{subequations}
\begin{align}
& H = H_{X} + H_{P} + H_{int} \label{aHam}\\
& H_{X} = \sum_{l,\alpha, n, {\vec q}} E_{X}^{(l)}( {\vec q} +
{\vec G}_n )
\beta^\dagger_{l,\alpha, {\vec q}+{\vec G}_n} \beta_{l,\alpha, {\vec
    q}+{\vec G}_n}, \label{Hx}\\
& H_{P} = \sum_{i,{\vec q}} \hbar\ome_{i,{\vec q}} a^\dagger_{i,{\vec q}} a_{i,{\vec
    q}} , \\
& H_{int} = \sum_{l,\alpha, n, i, {\vec q}} i\hbar
\ov{\Ome}_{l,\alpha, n, i, {\vec q}}
\beta^\dagger_{l,\alpha,{\vec q}+{\vec G}_n} a_{i,{\vec q}} + {\rm H.c.} .
\end{align}
\end{subequations}
The operator $\beta^\dagger_{l,\alpha, {\vec q}+{\vec G}_n}$ creates a $1s$
exciton in the $l$-th QW with polarization $\alpha=L,T$ (longitudinal or
transverse) and center of mass wavevector ${\vec q}+{\vec G}_n$ where 
${\vec G}_n=\frac{2\pi}{a}(n_x,n_y)$ is the 2D reciprocal lattice
vector for the photonic band structure with integers $n_x$ and $n_y$.
The exciton dispersion $E_{X}^{(l)}({\vec q})= E_{X}({\vec
  q}) + \delta E_l$ with $E_{X}({\vec q})\equiv E_{X0}
+ \frac{\hbar^2q^2}{2m_X}$ includes a fluctuation term $\delta E_l$
varying for different QWs to describe disorder effects.
$E_{X0}$ is the exciton recombination energy at zero center-of-mass
wavevector. $m_{X}=m_e+m_h$ is the 
effective mass of exciton with $m_e$ and $m_h$ being the electron and
hole effective masses, respectively. The operator 
$a^\dagger_{i,{\vec q}}$ creates a photon in the $i$-th band with
Bloch wavevector ${\vec q}$. The coupling matrix element is
calculated\cite{our} as:
\bea
\ov{\Ome}_{l,\alpha,  n, i, {\vec q}} &=& \frac{ |\phi(0)| d \sqrt{\ome_{i,{\vec q}} } }{
   S_{u.c.} \sqrt{2\hbar\vep_0}}  \nn\\
&& \mbox{} \times \int_{u.c.}
  d\vec{\rho}~ e^{-i{\vec G}_n\cdot {\vec \rho}} 
   u_{\alpha, i, {\vec q}} ({\vec
  \rho},z_l) \Theta({\vec
  \rho},z_l) . \label{ome-ov}
\eea
$|\phi(0)|$ denotes the amplitude of the $1s$ exciton wavefunction
when the distance between electron and  hole is zero. $d$ is the
inter-band dipole matrix element in CdTe. $z_l$ is the
coordinate of the center of the $l$-th QW in the $z$-direction.
$S_{u.c.}=a^2$ is the area of the unit cell in the $x$-$y$ 
plane. $u_{\alpha, i, {\vec q}}= {\vec {\bf e}}_{\alpha}\cdot 
{\vec u}_{i,{\vec q}}$, where ${\vec {\bf e}}_\alpha$ is an unit
vector along the polarization direction of the
$\alpha$ exciton. For longitudinal ($\alpha=L$) 
excitons ${\vec {\bf e}}_\alpha$ is along ${\vec q}$ while for
transverse ($\alpha=T$) exciton it is perpendicular to both ${\vec q}$
and the $z$ direction. $\Theta({\vec r})$ equals 1 when the coordinate
${\vec r}$ is in the semiconductor region and 0 when ${\vec r}$ is in
the air region. $\Theta({\vec r})$ describes the overlap between
the photonic field and the excitonic wavefunction. The integration 
in Eq.~(\ref{ome-ov}) is over the unit cell (u.c.) of the structure in
the $x$-$y$ plane.

It is sufficient to use only the spectrum close to the energy minima
of the lowest confined photonic band to calculate the dispersion of
the lower polariton branch (see Figs.~\ref{struc}d and \ref{disper}b). This branch is essential
for the study of polariton BEC. There are two energy minima for the
photon: one at ${\vec Q}^{(x)}=(\frac{\pi}{a}, 0)$, the other at
${\vec Q}^{(y)}=(0, \frac{\pi}{a})$. We consider the situation when
the exciton recombination energy $E_{X0}$ is close to the two energy
minima. The photonic dispersion around the two minima is approximately
parabolic,
\bea
\hbar\ome_{\vec q} &=& \hbar\ome_{0} + 
\frac{\hbar^2(q_x-Q_x^{(\nu)})^2 }{2m_x^{(\nu)}}
+
\frac{\hbar^2(q_y-Q_y^{(\nu)})^2}{2m_y^{(\nu)}} \label{ccav}
\eea
with $\nu=x,y$. From the $D_{2d}$ symmetry, the
photonic effective masses satisfy the relations $m_x^{(x)} = m_y^{(y)}$ and $m_x^{(y)} =
m_y^{(x)}$. If $a=360$~nm, plane wave expansion calculation\cite{mpb} yields 
$m_x^{(x)} = 2.9\times 10^{-5}m_0$ and $m_x^{(y)} = 4.7\times
10^{-5}m_0$ where $m_0$ is the electron mass in vacuum.
Scale invariance of the Maxwell equations dictates that $\ome_0$,
$m_x^{(x)}$, and $m_y^{(x)}$ are proportional to $1/a$.
For CdTe QWs, the effective mass of the conduction band electron is
$m_e=0.095m_0$\cite{madelung}, while the effective mass of the heavy-hole
is $m_h=0.4m_0$\cite{mhhp}.

When the fluctuation in the exction energy among QWs is neglected,
i.e., $E_{X}^{(l)}({\vec q})= E_{X}({\vec q})$ for all $l$, we
introduce the collective excitonic operator $b_{\vec q} \equiv \sum_{l, \alpha, n} \frac{
  \Ome_{l,\alpha, n, {\vec q}} }{\Ome_{\vec q} } \beta_{l,\alpha, n,
  {\vec q} }$ to obtain the effective polariton Hamiltonian\cite{yang1,yang2}: 
\begin{subequations}
\begin{align}
& H = H_0 + H_I , \\
& H_0 = \sum_{\vec q}\left[ E_{X}({\vec q}) b^\dagger_{\vec q} b_{ {\vec
      q}} + \hbar\ome_{\vec q} a_{\vec q}^\dagger
  a_{ \vec q} \right]  ,\\
& H_I = \sum_{\vec q}  i\hbar \Ome_{\vec q} (b^\dagger_{\vec q} a_{\vec
  q}-a^\dagger_{\vec q}b_{\vec q}) .
\end{align}
\label{heff}
\end{subequations}
The collective exciton-photon coupling is given by
\be
\hbar\Ome_{\vec q} = \hbar \sqrt{\sum_{l,\alpha, n} |\ov{\Ome}_{l,\alpha,
    n, {\vec q}}|^2} .
\ee
Here $b_{\vec q}$ corresponds to the mode that couples most strongly
to the lowest 2D guided mode in the PBG. Other excitonic modes 
orthogonal to this mode
(in the Hilbert space of excitonic states spanning all QWs) are much more weakly
coupled to the 2D photonic bands. In the above equations, the
photonic band index $i$ is omitted  since only the 
lowest confined 2D photonic band is relevant. We also approximate the
${\vec q}$ dependence of the collective coupling by its value at the
two polariton energy minima and define $\Ome\equiv
\hbar\Ome_{{\vec Q}^{(x)}}= \hbar\Ome_{{\vec Q}^{(y)}}$.
This is justified since beyond the narrow range around the two energy
minima, the lower polariton dispersion becomes exciton-like and its
detailed form does not alter the low energy dynamics (see
Figs.~\ref{struc}d and \ref{disper}b). The dispersion of the lower polariton branch is given by
\begin{align}
& E_{LP}({\vec q}) = \frac{E_{X}({\vec q})+\hbar\ome_{\vec q}}{2}
- \left[\left(\frac{E_{X}({\vec q})-\hbar\ome_{\vec
        q}}{2}\right)^2+\hbar^2\Ome^2\right]^{\frac{1}{2}} .\label{Epl} 
\end{align}
The above dispersion can likewise be extended to the whole 2D wavevector space 
since away from the energy minima ${\vec Q}^{(x)}$ and ${\vec
  Q}^{(y)}$, the dispersion of the lower polariton becomes
exciton-like (i.e., almost flat dispersion).

\section{Very strong light-matter interaction}

Fig.~\ref{vrs}a shows the collective exciton-photon coupling $\hbar\Ome$
calculated for different QW width when excitons are in resonance with
the lowest 2D guided mode for photons, i.e., $E_{X0}=\hbar\ome_0$. The vacuum Rabi splitting,
$2\hbar\Ome$, exceeds 100~meV when the QW width is 3~nm
(experimentally accessible as shown by Refs.~\cite{lz1,lz2}). The width of
the barrier layers between QWs is fixed at 5~nm throughout this work.
Fig.~\ref{vrs}b depicts how the photonic crystal lattice constant $a$ must be
adjusted as the QW width is varied, in order to maintain resonance
between the exciton recombination energy and the 2D guided mode. The
total number of QWs that fit into the active region is correspondingly
adjusted. Fig.~\ref{vrs}c depicts the variation of the exciton binding energy
and Bohr radius with QW width. The calculated
exciton binding energies, ranging from 17~meV to 23~meV, agree well with
experimental results of $17\sim 25$~meV\cite{Eb1,Eb2,sat-exp}.

Considerable enhancement of the exciton-photon coupling in our photonic
crystal microcavity (compared with Fabry-P\'erot microcavities)
originates from the light-trapping effect of the PBG. In Fig.~\ref{vrs}d we
compare the photonic field intensity $|{\vec {\cal E}}|^2$
distribution as a function of $z$, but averaged over the $x$-$y$ plane
for the Cd$_{0.4}$Mg$_{0.6}$Te/Cd$_{0.8}$Mn$_{0.2}$Te Fabry-P\'erot
and the 3D photonic crystal microcavities with 
the same photon energy. The photonic crystal 
microcavity clearly focuses the electric field much more strongly than the
Fabry-P\'erot. Moreover the averaged electric field intensity at $z=0$ is
significant larger. This is because in photonic crystal a large volume
fraction ($\simeq 75\%$) is air, which reduces the over all dielectric
constant of the structure. The reduced screening enhances the averaged
electric field intensity $|{\vec {\cal E}}|^2$ at $z=0$ by eight
times. Another important advantage of 
the PBG heterostructure are the strong spatial variations of the field
intensity in the $x$-$y$ plane. The higher intensity regions act as an
attractive ``optical potential'' for the exciton and the wavefunction
of the lower polariton is peaked in these ``hot spots''. This offers a
way to ``pattern'' the polariton condensates in the $x$-$y$ plane (see
Fig.~\ref{vrs}e). The peak intensity in the $x$-$y$ plane is greater than the
average intensity depicted in Fig.~\ref{vrs}d. This provides further
enhancement in exciton-photon coupling compared to the 1D
Fabry-P\'erot geometry for which the field intensity is uniform in the
$x$-$y$ plane. From Figs.~\ref{vrs}d and \ref{vrs}e the peak intensity is
about 38 times (the average intensity is 16 times) as large as that in
the Fabry-P\'erot microcavity. Other photonic crystals that have a
larger PBG, such as the slanted-pore photonic
crystals\cite{ovi}, provide even stronger light focusing and larger
exciton-photon couplings than we have presented here using the
woodpile architecture.

Another advantage of PBG light-trapping is that the quality factor of
the microcavity can be very high. A quality factor of 
$2\times 10^6$ with nanosecond photonic lifetime has been experimentally
demonstrated in the woodpile architecture\cite{high-q}. This quality
factor can be further enhanced by simply increasing the thickness of
the cladding photonic crystal. Our calculation indicates that moderate
fabrication imperfection does not reduce the quality factor
considerably due to the existence of the 3D PBG [see the end of next section].
When radiative decay is strongly inhibited, polaritons can decay through 
slower non-radiative processes. The Auger recombination
rate\cite{auger}  is estimated as about $10^{-3}$~second [see Appendix]. Non-radiative recombination triggered by electronic impurity
states such as the Shockley-Read-Hall mechanism\cite{srh}, may be more
important in polariton decay. Nevertheless, the measured
photoluminescence decay time of 150~ps in CdTe QW at room
temperature\cite{pl-decay} suggests that non-radiative polariton decay
time should be longer than 150~ps. The non-radiative exciton decay
time is believed to be on the order of nanoseconds\cite{rmp}. Clearly, very strong exciton-photon
coupling, long polariton lifetime, and small polariton effective mass
are possible in the photonic crystal microcavity.

\begin{figure}[]
  \centerline{\includegraphics[height=4.0cm]{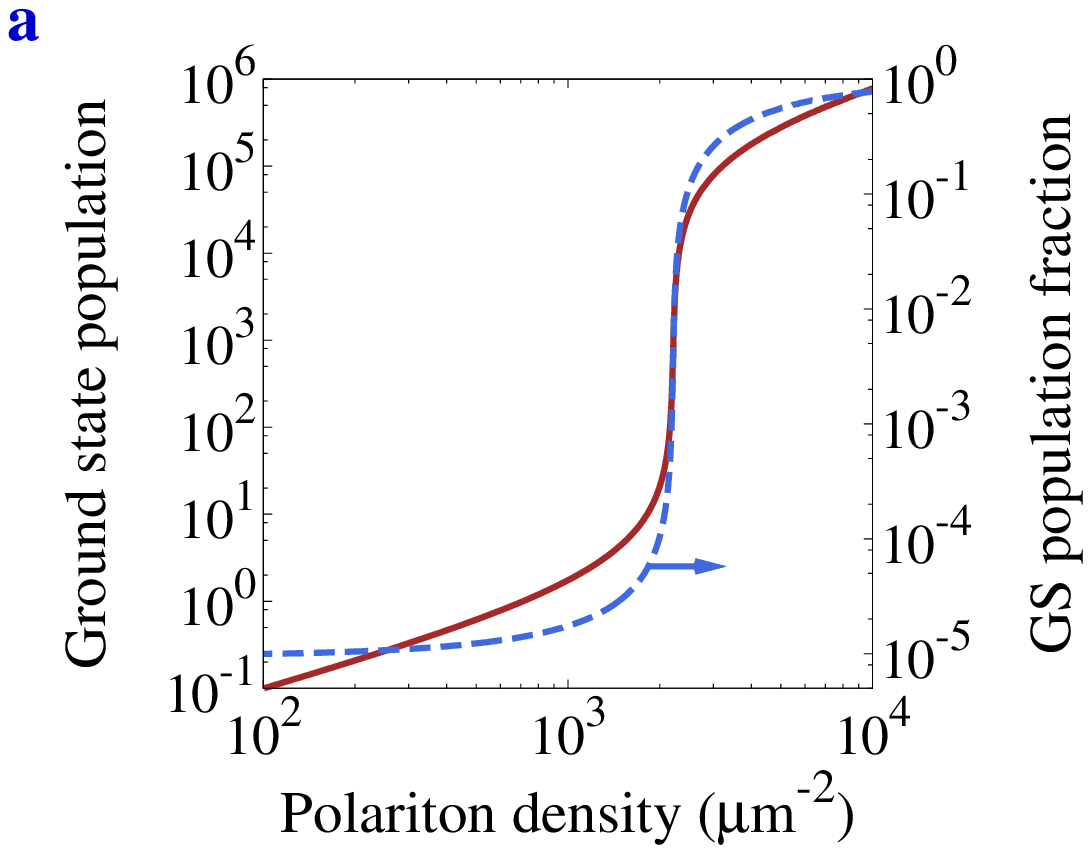}}
  \centerline{\includegraphics[height=4.0cm]{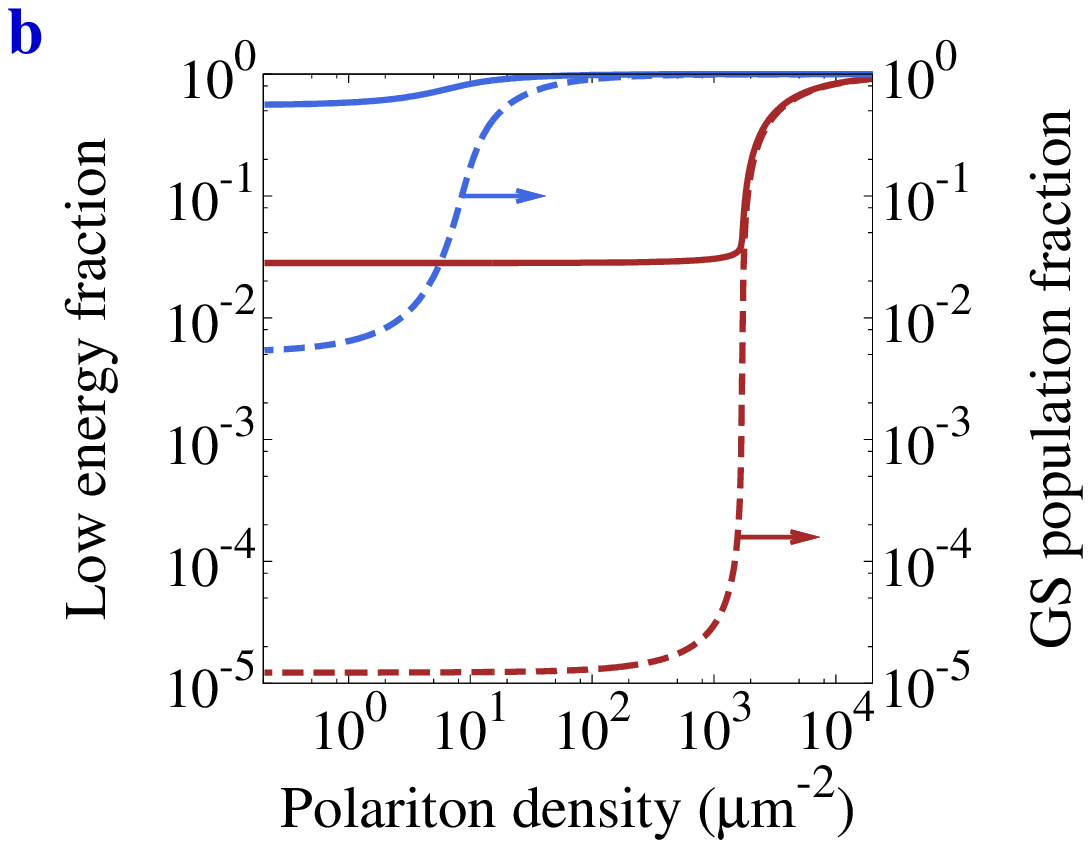}}
  \centerline{\includegraphics[height=3.8cm]{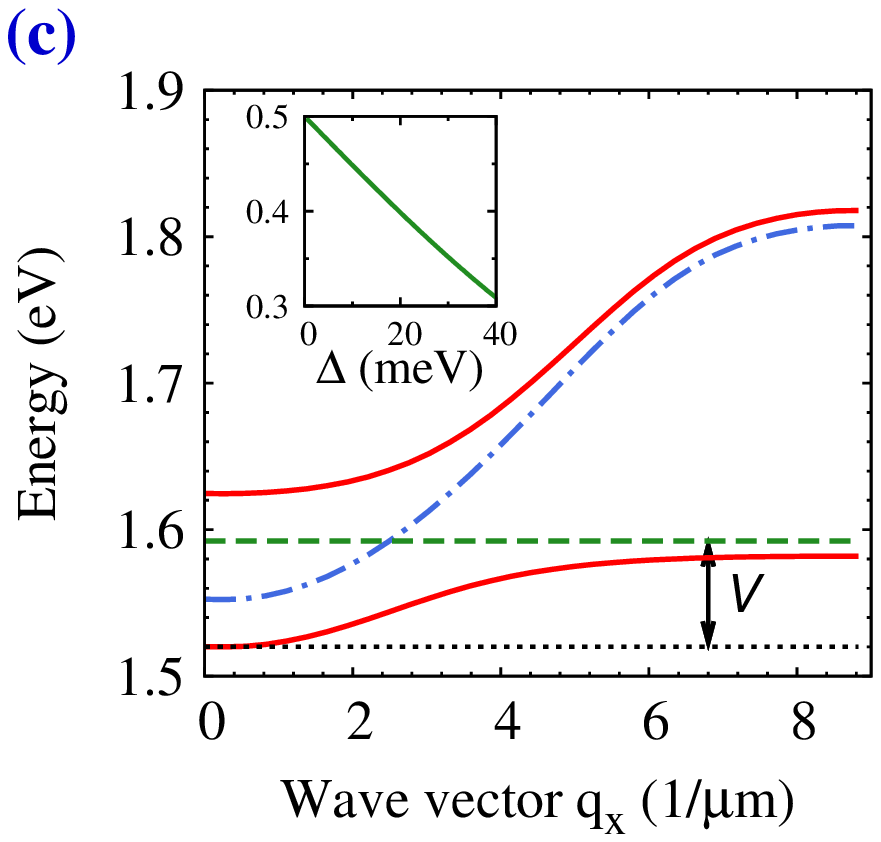}\includegraphics[height=3.8cm]{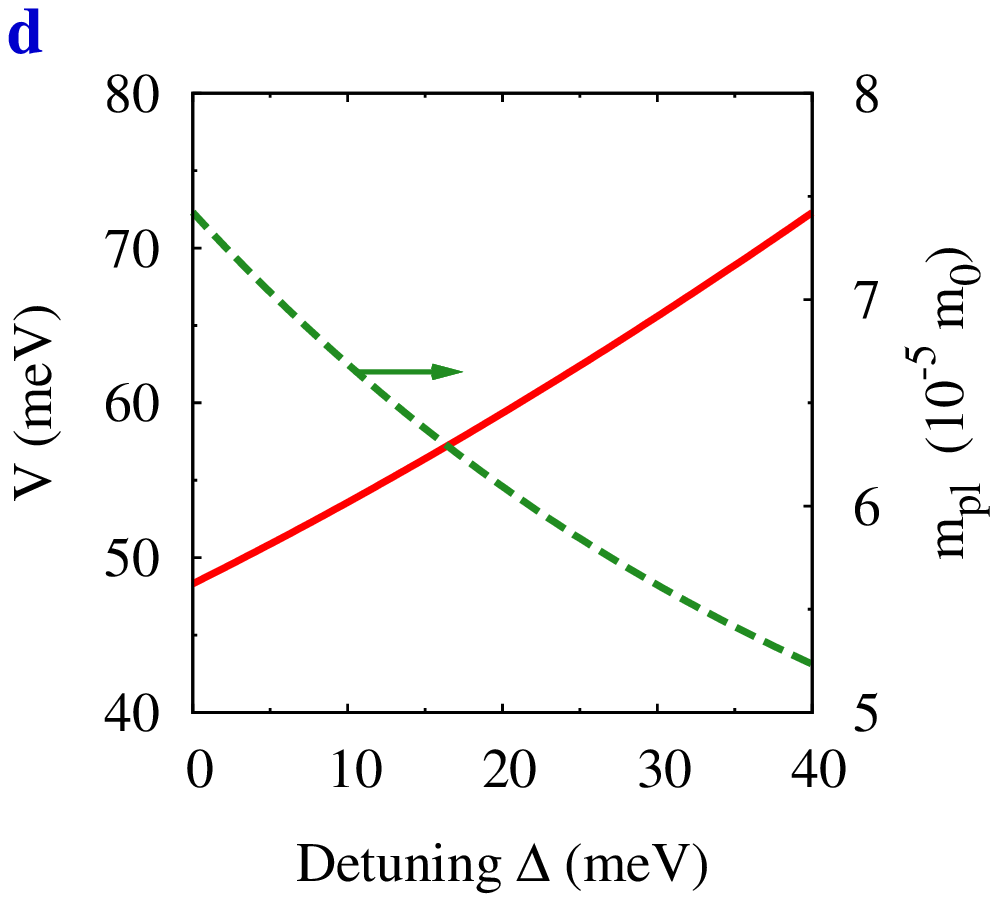}}
  \caption{ Polariton population on ground state and
      other low energy states for exciton box trap with
      $D=10~\mu$m, exciton-photon coupling of $\hbar\Ome=$48~meV,
      and exciton recombination energy of $E_{X0} = 1.59$~eV. The QW
    width is 4~nm and the barrier layer width is 5~nm. The detuning is
    $\Delta=30$~meV for (a) and $\Delta=40$~meV for (b).  (a) Population of the ground state (GS) 
    vs. polariton density (solid curve). The rapid, nonlinear increase
    of GS population indicates the BEC transition. The GS population
    fraction (dashed curve) i.e. GS population
    divided by the total population is of order unity for densities
    beyond the BEC transition. (b) GS population fraction (dashed
    curves) and the population fraction on the low energy discrete
    states (solid curves) below the continuum exciton recombination
    energy $E_{X0}$ (there are more than 1000 such states). These are
    photon-like states with very small effective mass, whereas states
    higher than $E_{X0}$ are exciton-like. When a large
    population fraction resides in the photon-like states, the BEC threshold
    density is reduced. For $T=100$~K (blue curves), populated
    polariton states are mostly photon-like. The threshold density is
    much lower than for $T=300$~K (brown curves) where the population
    fraction on low energy states is negligible. (c) 
    The dispersion of polaritons (red solid curves), bare exciton
    (green dashed curve), and pure photon (blue chain) for detuning
    $\Delta=40$~meV. The dispersion depth of 
    the lower polariton branch $V$ is also depicted (the black dotted line
    denotes the lower polariton band edge). Inset: excitonic fraction
    of the lower polariton branch as a function of detuning $\Delta$.
    (d) The dispersion depth $V$ (red solid curve) and
    polariton effective mass $m_{pl}$ (green dashed curve)
    of the lower polariton branch vs. the detuning $\Delta$.}
  \label{pop}
\end{figure}

\section{Room temperature polariton BEC}

In this section we delineate the precise conditions for room
temperature polariton BEC in our photonic crystal microcavity. The BEC
transition temperature is very sensitive to the dispersion depth of
the lower polariton branch and the density of polaritons, but
relatively insensitive to the polariton trapping area over a broad
range of lengths, $D$, from 10~$\mu$m to 1~cm, when the polariton density
is high. The polariton density can be controlled by excitation of
electron-hole pairs with energy above the upper 3D photonic band
edge. These carriers can then relax into the lower polariton branch
within the PBG. Since the polariton lifetime is enhanced beyond the
thermalization time (on the order of ps at room temperature),
polaritons achieve thermal equilibrium long before they decay. We
consider a finite-sized trap with box (square well potential)
confinement in the $x$-$y$ plane. This square quantum box has a side
length of $D$. Such a box can be created by replacing our infinite
central slab by a finite-sized slab\cite{high-q,noda,multi2} or by
increasing the Mg fraction outside the prescribed box
region\cite{trapping} to create a barrier for excitons. Trapping by
these means enables higher BEC transition temperature than 
harmonic trapping potentials induced by strain\cite{yang3}. The trapping
box induces quantization of the polariton wavevector in $x$ and $y$ directions,
discretizing the polariton spectra, and enabling the BEC phase
transition that would otherwise be forbidden in an infinite 2D system. To simplify the calculation we approximate
the polariton effective mass with the isotropic ``density-of-states mass'' $m_{dos} = \sqrt{m_x^{(x)}m_x^{(y)}}$.

In Figs.~\ref{pop}a and \ref{pop}b, we plot the ground state occupation $N_0$ as a function
of the polariton density. The polariton distribution is
calculated assuming a non-interacting, equilibrium polariton
gas. There is a rapid increase of $N_0$ around a threshold density,
above which $N_0$ becomes comparable with the total polariton number
$N_{tot}$. We define the polariton BEC transition according to the
criterion $N_0/N_{tot}=0.1$\cite{onsager} This criterion
gives lower estimate of the BEC transition temperature than the
criterion employed by Ketterle and van Druten\cite{ketterlepra}. 
At low temperatures, polaritons are mainly distributed in the low energy,
photon-like states around the bottom of the lower polariton branch
with very small effective mass (see Fig.~\ref{pop}c). The small
effective mass enhances condensation and reduces the threshold density
for polariton BEC. The dispersion depth of these photon-like states is
\be
V \equiv \left. (E_{X} - E_{LP})\right|_{{\vec q}=0} =
\frac{\Delta}{2}  + \sqrt{\left(\frac{\Delta}{2}\right)^2+\hbar^2\Ome^2}
, \label{Pi}
\ee
where $\Delta = E_{X0}-\hbar\ome_0$ is the detuning between the
exciton and photon. On the other hand, if the temperature is
comparable or higher than $V/k_B$, the occupation of high energy, exciton-like
states is significant. The effective mass of the bare exciton is
around four orders of magnitude larger than the 2D guided photon
effective mass. Likewise the density of exciton-like states is about
four orders of magnitude larger than the low-energy, photon-like
states. The consequences of this are indicated in Figs.~\ref{pop}a (low
temperature case) and \ref{pop}b (high temperature case). Polariton BEC at
temperatures comparable with $V/k_B$ require very high polariton
density. Since the polariton is a quantum superposition of a photon
and an exciton, the polariton density cannot exceed a specific
``saturation density'', beyond which exciton-exciton interaction cause
considerable exciton ionization. Therefore, the polariton BEC
transition temperature is essentially limited by the dispersion depth
$V$. The dispersion depth is increased by a positive detuning
$\Delta$ (see Fig.~\ref{pop}d). However, the excitonic fraction of the polariton is reduced
with increasing $\Delta$. This reduces the 
phonon-polariton and polariton-polariton interactions and consequently
it takes longer time for polaritons to reach their thermal
equilibrium. From experimental measurements\cite{therm} the
phonon-polariton scattering time at room temperature is estimated to
be about 0.1~ps [see Appendix]. A thermalization time on the
order of 1~ps is possible if the exciton fraction is higher than
10\%. In the range of detuning considered in this article,
$0<\Delta<40$~meV, the excitonic fraction is greater than 19\% for
exciton-photon coupling as low as $\hbar\Ome=25$~meV. Both the
excitonic fraction and the effective mass of the lower
polariton dispersion minimum decrease with detuning $\Delta$ as shown
in the inset of Fig.~\ref{pop}c and in Fig.~\ref{pop}d, respectively.

\begin{figure}[]
  \centerline{\includegraphics[height=3.8cm]{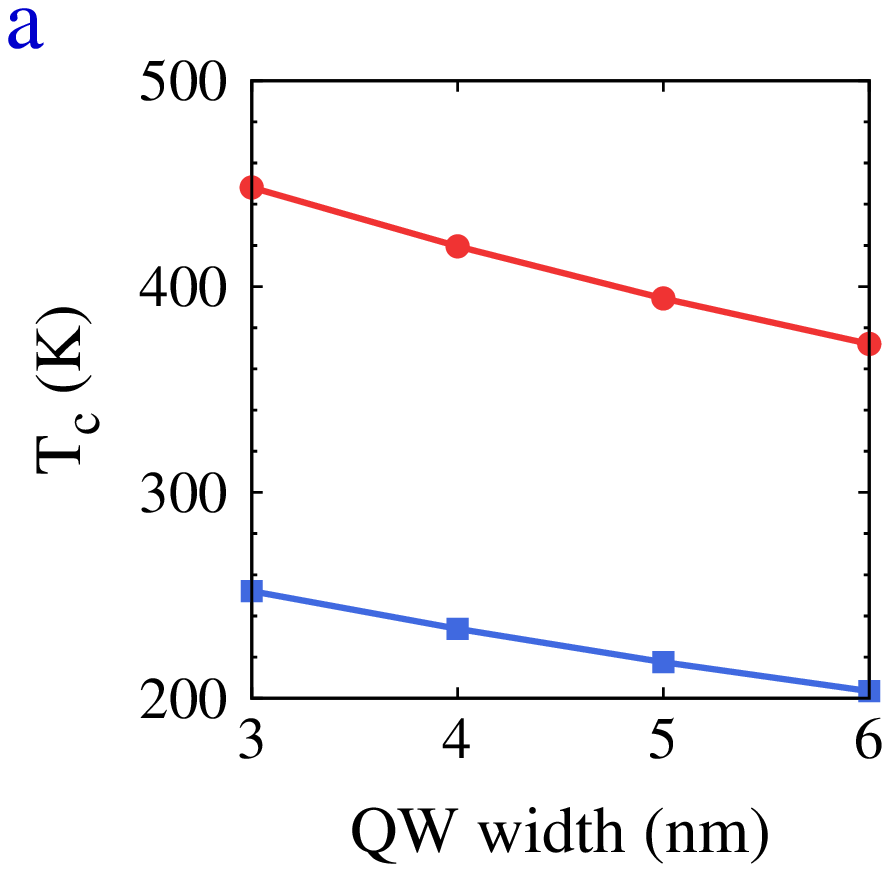}}
\centerline{\includegraphics[height=3.7cm]{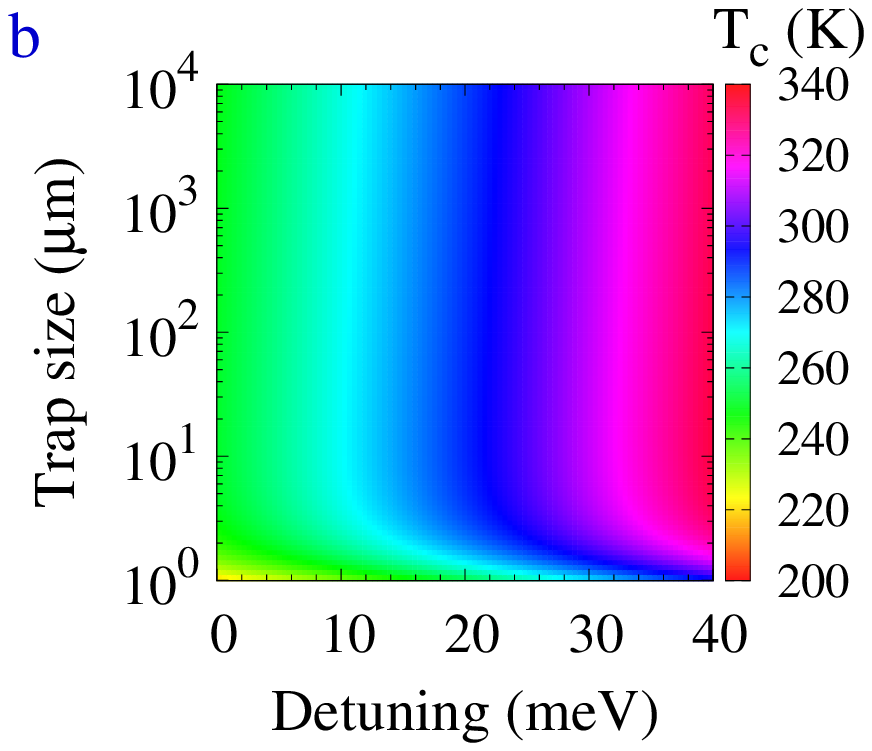}\includegraphics[height=3.7cm]{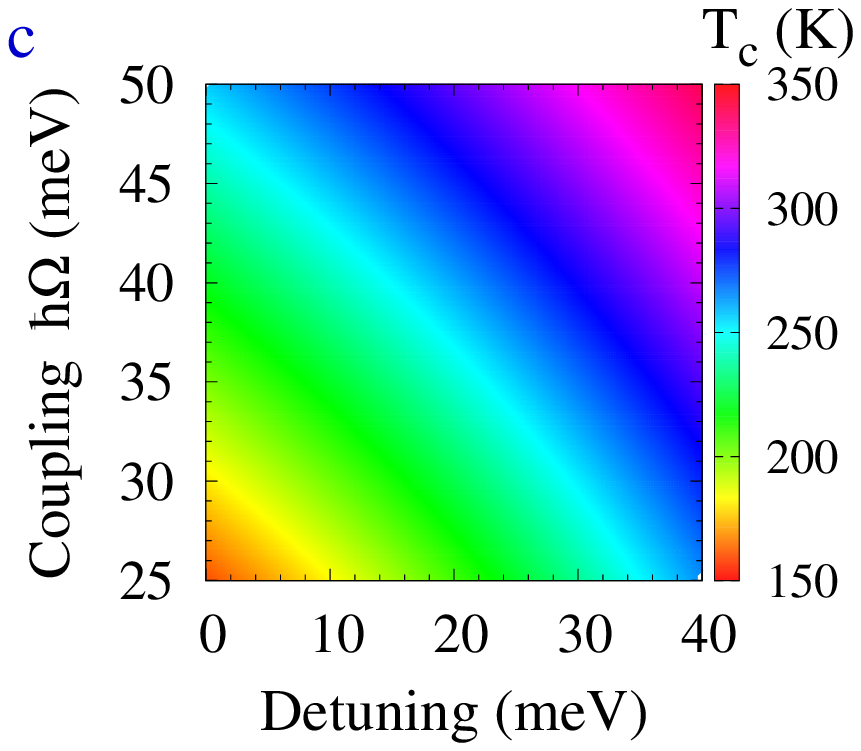}}
  \caption{ Polariton BEC transition temperature $T_c$
    at fixed densities. (a) Transition temperature $T_c$ with
    detuning $\Delta=30$~meV for different QW
    width (with QWs filling entire active region) at fixed polariton densities: $1.5\times 10^4~\mu{\rm m}^{-2}$ (curve
    with solid circles) and $1.9\times 10^3~\mu{\rm m}^{-2}$ (curve
    with solid
    squares). For the higher total density,
    the exciton density in each QW is below $(9a_B)^{-2}$. For the
    lower total density, the exciton density in each QW is
    below $(26a_B)^{-2}$. (b) Dependence of the
    transition temperature $T_c$ on the detuning $\Delta$ and the
    exciton trap size $D$ when the exciton-photon coupling is
    $\hbar\Ome=48$~meV (QW width 
    4~nm and barrier width 5~nm) and the polariton density is 
    $3\times 10^3~\mu{\rm m}^{-2}$. (c) Dependence of the
    transition temperature $T_c$ on the detuning $\Delta$ and the
    exciton-photon coupling $\hbar\Ome$ when the exciton trap size 
    is $D=10~\mu {\rm m}$ and the polariton density is $3\times
    10^3~\mu{\rm m}^{-2}$.}
  \label{tc}
\end{figure}

We calculate the polariton BEC transition temperature $T_c$ for
various QW widths, exciton trap sizes, detunings, and exciton-photon
couplings for fixed polariton density. The results are plotted in
Fig.~\ref{tc}. Clearly, room temperature polariton BEC is attainable
at overall polariton density $\gtrsim 3\times 10^3~\mu{\rm
  m}^{-2}$. This is well below the saturation density in each QW of
$(5a_B)^{-2}$ where $a_B$ is the exciton Bohr radius. The
transition temperature is higher for smaller QW width since a larger
number of QWs can be used. It is likewise higher for larger
exciton-photon coupling and positive detuning. This is consistent with the
picture that $T_c$ is limited by the dispersion depth
$V$. For large trap sizes $D\gg 10~\mu$m, $T_c$ decreases as expected
from the Mermin-Wagner theorem\cite{wagner} that forbids finite
temperature BEC in an infinite 2D system. However, this decrease is
very slow and little degradation in $T_c$ is seen over the range of
$10~\mu{\rm m}<D<1$~cm. Interestingly the exciton trap size dependence
of $T_c$ for smaller $D$ is nonmonotonic: $T_c$ has a peak value
around a box size $D\simeq 5~\mu$m. When the exciton trap size is
sufficiently small, the quantization energy of the cavity (e.g., the energy difference between the first
excited state and the ground state) can be comparable or larger than
the dispersion depth $V$. The condensation then crosses over from
polariton-like to exciton-like. Since the exciton effective mass is
much larger than that of polariton, the transition
temperature is rapidly reduced.

The polariton density used in our calculation is experimentally
accessible. For MQWs the total exciton density is distributed among
the QWs and the excitonic fraction in each QW of a polariton may be
much less than unity. For example, using $N=100$ QWs, each of width
4~nm and barrier width 5~nm (i.e., $\hbar\ome_0=1.56$~eV and
$E_{X0}=1.59$~eV with $a=365$~nm and $a_B=3.8$~nm), the collective
exciton-photon coupling is 48~meV. For a detuning of $\Delta=30$~meV,
the largest excitonic fraction in any single QW is 5.5\%
[the exciton fraction varies among  QWs since the
exciton-photon coupling for each QW is different, see
Eq.~(\ref{ome-ov})]. For total polariton density $1.5\times 10^4~\mu{\rm
  m}^{-2}$, the exciton density in each QW is no larger than
$(9a_B)^{-2}$ where $a_B$ is the exciton Bohr radius. This single QW
exciton density is less than the saturation
density\cite{sat-exp,saturation} $\simeq (5a_B)^{-2}$.

\begin{figure}[]
  \centerline{\includegraphics[height=3.9cm]{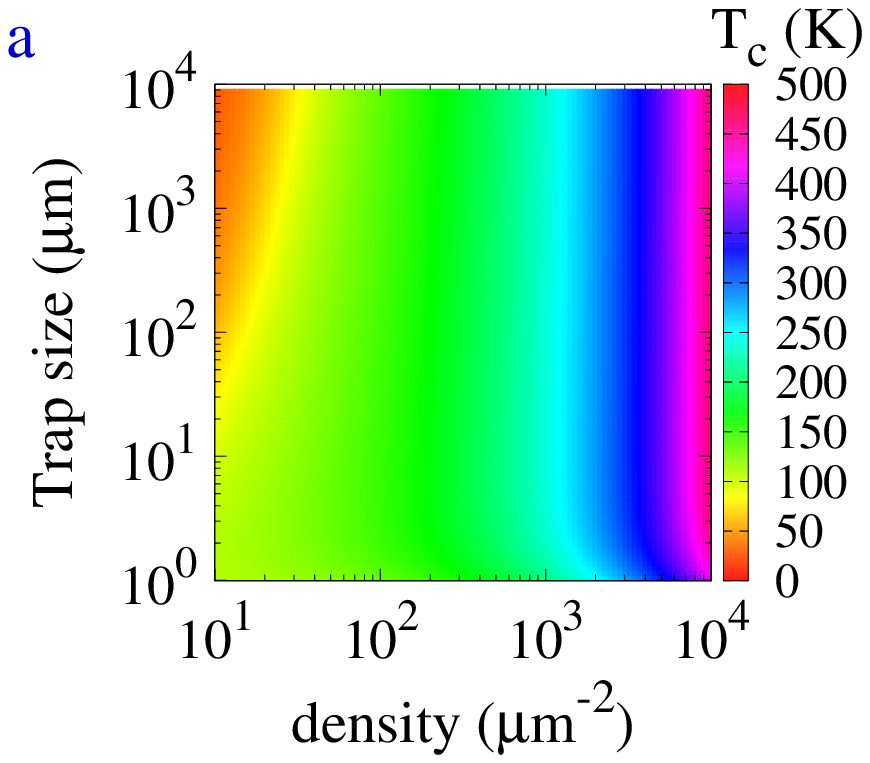}\includegraphics[height=3.9cm]{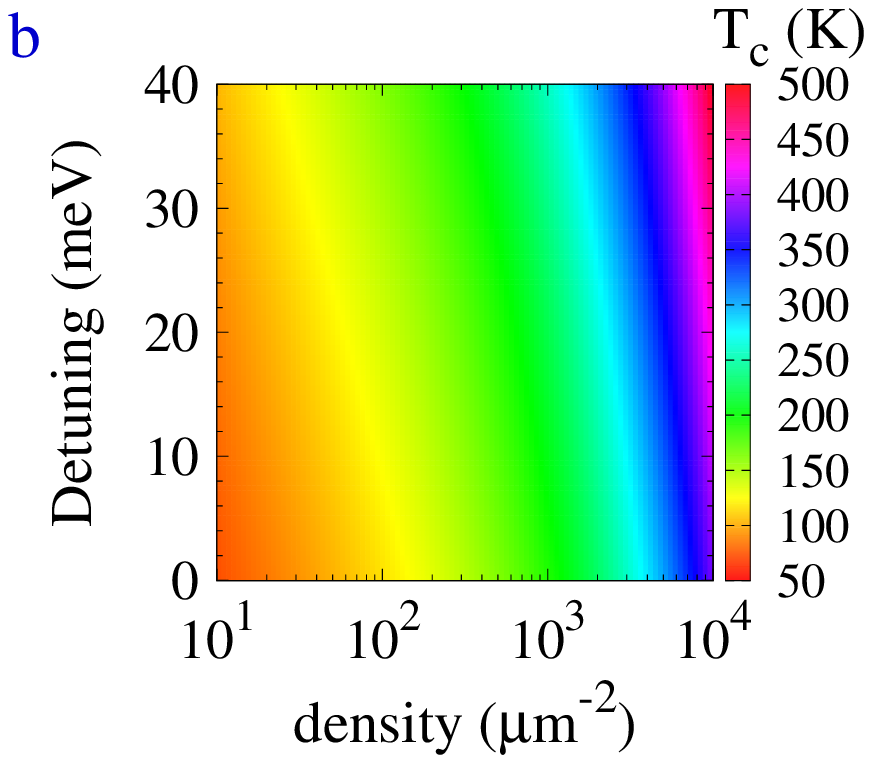}}
  \centerline{\includegraphics[height=3.9cm]{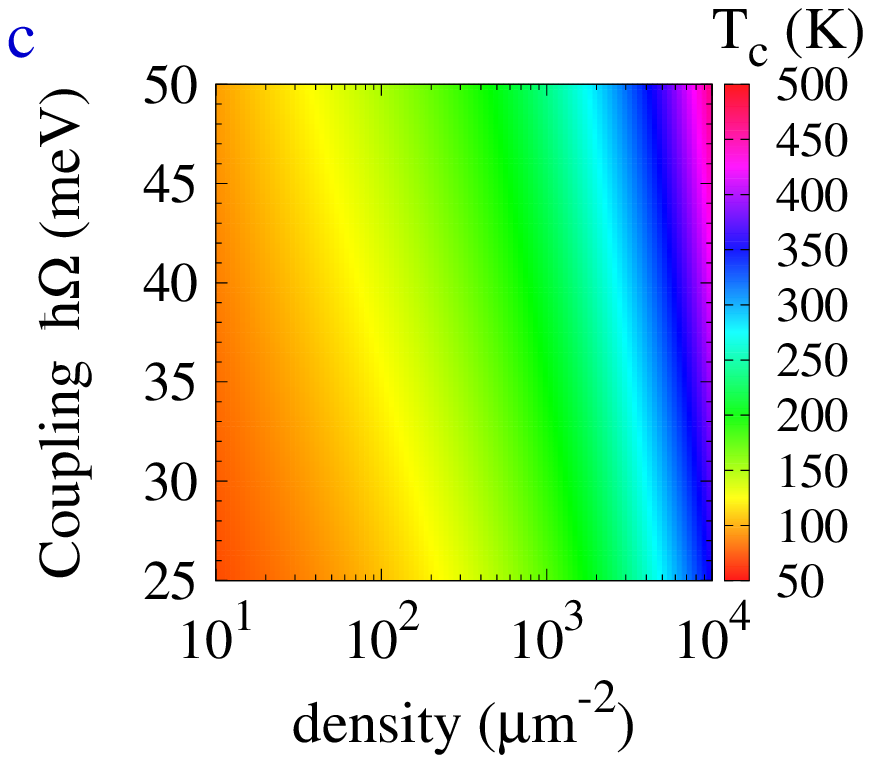}\includegraphics[height=3.9cm]{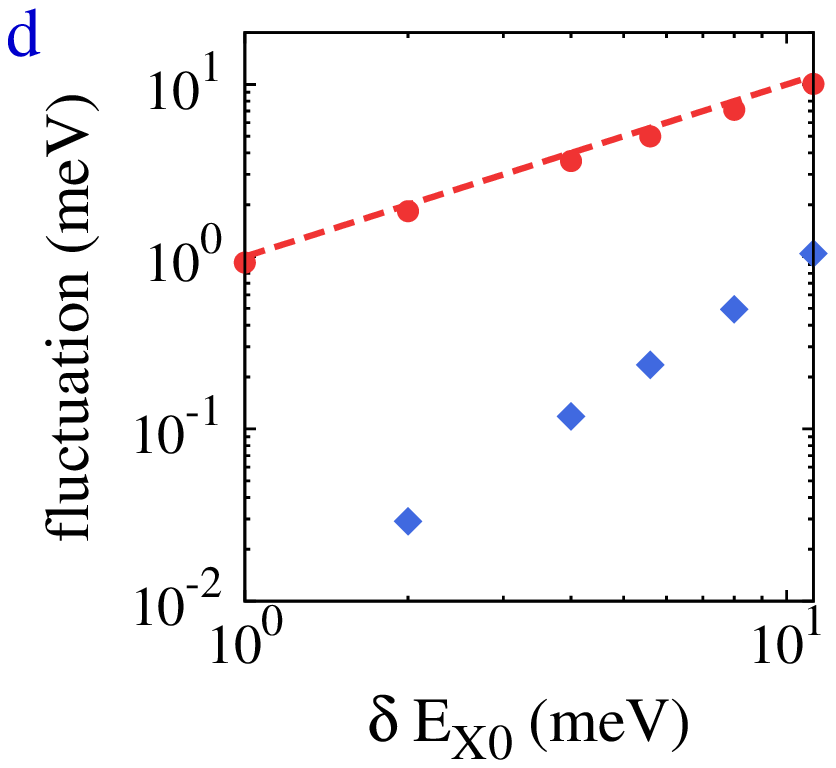}}
  \caption{ BEC transition temperature $T_c$ as a function of total
    polariton density and (a) box trap size, (b) exciton-photon
    detuning, and (c) exciton-photon coupling strength. (d) Root mean
    square deviation of the vacuum Rabi splitting (diamonds) and lower
    polariton dispersion depth (circles) as a function of exciton
    inhomogeneous broadening $\delta E_{X0}$. In (a) and (b) the
    exciton-photon coupling is $\hbar\Ome=48$~meV (QW width 4~nm and
    barrier width 5~nm). For (a) and (c) the detuning is 30~meV. For
    (b) and (c) the exciton box trap size is 10~$\mu$m. The exciton
    recombination energy is $E_{X0} = 1.59$~eV. For (d), the detuning
    is $\Delta = 0$. The dashed curve in (d) depicts the value of
    $\delta E_{X0}$ itself for reference.}
  \label{tc-n}
\end{figure}

Systematic dependence of the BEC transition temperature on total polariton
density is shown in Fig.~\ref{tc-n}. Fig.~\ref{tc-n}a  reveals  that
the exciton trap size dependence of $T_c$ is much more pronounced at low
polariton density where the transition temperature
is much smaller than $V/k_B$. Here most polaritons are low energy
photon-like states with very small effective mass. The transition
temperature at fixed density is limited mainly by the level spacing
between the ground and first excited state\cite{ketterlepra,yang3},
which depends sensitively on the exciton trap size. In contrast, at high polariton density, the transition
temperature is close to $V/k_B$. When the dispersion depth $V$ is the
main factor limiting $T_c$, the exciton trap size has marginal
influence. Figs.~\ref{tc-n}b and \ref{tc-n}c reveal that 
large detuning and/or exciton-photon coupling is favorable for room
temperature polariton BEC because both of them enhance the dispersion depth $V$
(see Fig.~\ref{pop}d).

Finally we discuss the effect of fabrication disorder. This can
randomize sensitive parameters such a s the detuning $\Delta$ and the
collective exciton-photon coupling $\hbar\Ome$. We show that the
former effect is more significant. Photonic disorder arises from
fluctuation of the shape and size of the rods and slabs constituting
the photonic crystal. Electronic disorder arises from impurities,
dislocations, (local) QW width fluctuation, random strains, etc. In addition
to these forms of inhomogeneous broadening, dynamic fluctuations
(homogeneous broadening) arise from phonon-polariton and
polariton-polariton interactions. These are particularly important at
high temperatures. 

Plane wave expansion calculation indicates that the 2D photonic band edge varies
by less than 2\% of its value if the fabrication error is smaller than
20~nm for woodpile rods outside the active region, less
than 12~nm for rods in the active woodpile region, and less than 6~nm
for the central slab thickness. The effect of electronic disorder has
been studied earlier\cite{yang2} for the single QW case, where it was found that
polaritons remain robust due to their small effective mass and highly
mobile half-photon nature. Electronic scale disorder is largely
averaged out in a scenario similar to motional narrowing\cite{yang2}.

For the MQW structures studied here, there are
also fluctuations of exciton energy among QWs due to variations in QW
width or lateral confinement of the rods. This broadening is modeled by adding a
random energy $\delta E_l$ to the exciton dispersion in the $l$-th QW,
$E_{X}^{(l)}({\vec q}) 
= E_{X0} + \frac{\hbar^2q^2}{2m_X} + \delta E_l$. 
$\delta E_l$ is modeled as a Gaussian random
variable with variance $(\delta E_{X0})^2$. The effects of this
disorder on vacuum Rabi splitting $2\hbar\Ome$ and polariton
dispersion depth $V$ are presented in Fig.~\ref{tc-n}d. The
calculation is performed by averaging over $10^5$ random
configurations. The standard deviation of the vacuum Rabi splitting is very small due to
the collective nature of the polariton. The fluctuation in polariton
energy consists of sum of random exciton energy variations from each
of the $N$ QWs. The total effect is an average effect which has a much
smaller variance according to the central limit theorem\cite{our},
$\frac{\delta\Ome}{\Ome}\sim {\cal O}\left(\frac{1}{\sqrt{N}}\right)$. 

The fluctuation of the polariton dispersion depth $V$ is approximately
the same as the inhomogeneous broadening of the exciton, $\delta E_{X0}$. The
randomization of the detuning $\Delta$ 
directly affects this dispersion depth [see Eq.~(\ref{Pi})]. This effect
is much more significant than the fluctuation of the vacuum Rabi
splitting $2\hbar\Ome$.
Experimental measurements in a single CdTe/CdMgTe QW give a typical
inhomogeneous broadening of 1~meV\cite{exp-dEx}. Hence intra-QW
inhomogeneous broadening can be rendered small with present-day
fabrication capabilities. Inter-QW inhomogeneous broadening estimated
from lateral confinement difference for QWs in rods and in the central
slab is less than 1~meV as well.

Finally, dynamic fluctuations (such as polariton-phonon and
polariton-polariton scattering) cause both homogeneous broadening and
relaxation of polaritons. The latter is crucial for polaritons
to reach the ground state and achieve thermal equilibrium
BEC. However, homogeneous broadening reduces the
polariton dispersion depth $V$ and consequently $T_c$. In CdTe QWs, at
room temperature, experiments\cite{therm} suggest a homogeneous linewidth of
$\Gamma=22$~meV [see Appendix]. The broadening of the cavity photon is
negligible for high quality factor ($2\times 10^6$ as shown in
Ref.~\cite{high-q}) photonic crystal microcavities. If we estimate the
effect of homogeneous broadening by adding an imaginary part,
$i\Gamma$, to the exciton energy, the VRS at zero detuning
($\Delta=0$) is reduced 
to $2\sqrt{\hbar^2\Ome^2 -\Gamma^2/4}$\cite{qw-broad}. More
importantly, the dispersion depth of the lower polariton for nonzero
detuning becomes:
\be
V= \frac{\Delta}{2}  +
\Re\sqrt{\left(\frac{\Delta-i\Gamma}{2}\right)^2+\hbar^2\Ome^2}
\ee
In general, exciton-phonon interaction reduces the dispersion depth
$V$. However, for $\Delta=30$~meV, $\hbar\Ome=48$~meV, and
$\Gamma=22$~meV, the lower polariton dispersion depth is reduced by
only 1.1~meV. This is equivalent to a reduction of detuning by
2~meV and has negligible effects on the transition temperature
according to Fig.~\ref{tc-n}b. 
These results
suggest that our very strongly coupled exciton-photon system can
withstand considerable homogeneous and inhomogeneous broadening and
still deliver equilibrium room temperature polariton BEC.

\section{Conclusion}

In summary, we have identified a CdTe-based photonic band gap
architecture with a range of different quantum well embeddings that
are predicted to support room temperature (or higher) equilibrium
Bose-Einstein condensation of exciton-polaritons. The most crucial
factors in enabling high temperature BEC are the dispersion depth of
the lower polariton branch and the total areal density of long-lived
excitons within the trapping region. The dispersion depth is
determined by the collective exciton-photon coupling strength and is
enhanced by detuning the bare exciton recombination energy above the
lower edge of a slow-light, 2D guided photonic band within the larger
3D photonic band gap. The 3D PBG enables considerably stronger light
trapping than conventional 1D Fabry-P\'erot microcavities, leading to
stronger exciton-photon coupling. The collective exciton-phonon
coupling strength is further enhanced by embedding up to about 100
quantum wells in the photonic crystal regions adjacent to a central
slab region containing two principal quantum wells. Positive detuning
of the exciton recombination energy relative to the strongly-coupled
photon mode is made possible by the 3D PBG. The positively detuned
exciton remains within the 3D PBG and it cannot decay radiatively into
extraneous optical modes that leave the system.

The total areal density of excitons is enhanced by distributing them
over a large number of quantum wells that are all strongly coupled to
the 2D guided mode within the 3D PBG. The use of CdTe-based
architectures offers a suitable balance between photonic and
electronic properties. The relatively low dielectric constant of CdTe
compared to other popular materials (such as GaAs) enables a small
exciton Bohr radius and higher saturation exciton
density. Nevertheless, the dielectric constant is sufficiently large
to facilitate a 3D PBG that encompasses the lower polariton dispersion
depth and provides strong 3D light confinement.

Our calculations of the BEC transition temperature are based on a
non-interacting polariton gas confined in a 2D box trap. Previous
studies\cite{yang3} have shown that a slight reduction in the BEC
transition, together with polariton anti-bunching effects are expected
when exciton-exciton interactions are included in the microscopic
Hamiltonian. While a box trap of 5~$\mu$m side length is predicted to
yield the highest BEC transition temperature, very little degradation
of $T_c$ is seen for box dimensions up to 1~cm when polariton density
is high. This suggests that BEC may be observable in a finite-size
photonic crystal up to centimeter-scale lateral dimension, without
recourse to a separately engineered 2D trap.

It is significant that for total polariton densities approaching
$10^4$~$\mu$m$^{-2}$, the exciton density per quantum well remains
below $(10a_B)^{-2}$, where $a_B$ is the exciton Bohr radius. At these
densities, our PBG architecture provides sufficiently strong
exciton-photon coupling and polariton dispersion depth to provide a
predicted BEC transition as high as 500~K, in the absence of
additional homogeneous and inhomogeneous broadening effects. This
offers a safe margin for the realization of room temperature BEC in
realistic systems with both foreseen and unforeseen fabrication
imperfections, exciton-phonon scattering processes and other damping
effects. In this way, the photonic band gap architecture provides a
realistic prospect of bringing the magic of  macroscopic quantum
mechanics to the realm of everyday experience.

\section*{Acknowledgments}
This work was supported by the Natural Sciences and Engineering
Research Council of Canada, the Canadian Institute for Advanced
Research, and the United States Department of Energy Contract
No. DE-FG02-10ER46754.

\begin{appendix}

\section{Calculation Methods and Details}

The spectrum and wavefunction of the photonic bands are calculated
by solving the Maxwell equations using the plane wave expansion
method\cite{mpb}. In calculating the photonic bands in
Cd$_{0.4}$Mg$_{0.6}$Te/Cd$_{0.8}$Mn$_{0.2}$Te Fabry-P\'erot cavity the
dielectric constants for Cd$_{0.4}$Mg$_{0.6}$Te is 7.0 (linear
interpolation between the dielectric constant of CdTe,
8.5\cite{dielectric}, and that of MgTe, 6.0\cite{mgte-diel}) while the
dielectric constant of Cd$_{0.8}$Mn$_{0.2}$Te is 7.8\cite{cdmnte}.
The electronic structures are calculated using the effective mass
approximation\cite{harrison}. The effective mass of the electron is
$m_e=0.095m_0$ with $m_0$ being the bare electron mass
in vacuum\cite{madelung}. For the [001]-grown QWs concerned in this work, the
heavy-hole effective mass along the growth direction is 
$0.6m_0$\cite{mhhz}, while the effective mass in the QW
plane is $m_h=0.4m_0$\cite{mhhp}. The energy and wavefunctions of the
electron and hole subbands are calculated using a finite-depth well
model\cite{harrison}. The electronic band gap of CdTe is $E_g=$1.475~eV\cite{madelung,band-gap},
while the electronic band gap for MgTe is 3.49~eV\cite{mgte-Eg}. Linear
interpolation yields the band gap of Cd$_{0.6}$Mg$_{0.4}$Te as
2.28~eV. The band gap difference between Cd$_{0.6}$Mg$_{0.4}$Te and
CdTe is 0.8~eV. The offset for the conduction band between those two
materials is 70\% of their band gap difference\cite{Ecv}. The
interband dipole matrix element is given by
\be
d =  \frac{e \hbar \sqrt{E_{cv}} }{ E_g\sqrt{ 2 m_0} } ,
\ee
where $E_{cv}=2 P_{cv}^2 /m_0$ is the energy related to the momentum
matrix element, $P_{cv}$, between the conduction and valence bands. From
Ref.~\cite{Ecv} we take $E_{cv} = 21$~eV. Material parameters in the
above are taken for room temperature.

The exciton spectrum and wavefunctions are calculated by
diagonalizing the exciton Hamiltonian. For our narrow
QWs, we consider only the excitons formed by electrons and holes in
their lowest subband\cite{Eb2}. We ignore the mixing with higher
subbands and diagonalize the exciton Hamiltonian within the lowest
subband numerically. This calculation yields an exciton binding energy
comparable to the experimentally observed one. The exciton wave
amplitude $|\phi(0)|$ and exciton Bohr radius are also obtained, using
the same considerations.

The Auger recombination time, $\tau_A$, is estimated using the method
presented in Ref.~\cite{auger}. For a volume exciton density of 
$10^{17}{\rm cm}^{-3}$ (volume density of excitons corresponding to an
areal polariton density of $10^{4}\mu{\rm m}^{-2}$,
estimated by dividing the areal polariton density by the
thickness of the active region) in CdTe, $\tau_A\simeq
10^{-3}$~second. This very small Auger recombination rate is due to
the large band gap and small carrier density. Another
non-radiative decay mechanisms is the Shockley-Read-Hall mechanism
due to electronic trapping centers within the semiconductor band
gap. However, experiments in high quality CdTe QWs have shown
photoluminescence decay time of 150~ps at room
temperature\cite{pl-decay}. This places a lower bound on the
non-radiative decay time.

The polariton thermalization time is estimated as
follows. Experiments\cite{therm} have found that exciton homogeneous
broadening at high temperature is dominated by interactions with
longitudinal-optical (LO) phonons. The exciton line width
is given by $A_{LO}/[\exp(E_{LO}/k_BT)-1]$ where $E_{LO}=21.3$~meV is
the LO-phonon energy\cite{madelung} and the parameter $A_{LO}=15$~meV
is fitted from the measured temperature dependence of the exciton line
width in Ref.~\cite{therm}. Using this result, we estimate the
exciton line width due to LO-phonon scattering at $T=300$~K is
22~meV. This suggests a polariton line width at zero detuning 
($\Delta=0$) of 11~meV (exciton fraction of polariton is 50\%) and
corresponds to a sub-picosecond thermalization time. For positive
detuning, if the exciton fraction in the polariton remains larger than
10\%, then the thermalization time is below or around 1~picosecond. This is
considerably smaller than the non-radiative decay time of polariton, and suggests
that the polariton gas has sufficient time to reach thermal
equilibrium.

\end{appendix}


\begin{thebibliography}{99}


\bibitem{weisbuch1} C. Weisbuch, M. Nishioka, A. Ishikawa, and
  Y. Arakawa, {\it Observation of the coupled exciton-photon mode
  splitting in a semiconductor quantum microcavity},
  Phys. Rev. Lett. {\bf 69}, 3314 (1992).

\bibitem{koch} G. Khitrova, H. M. Gibbs, M. Kira, S. W. Koch, and
  A. Scherer, {\it Vacuum Rabi splitting in semiconductors}, Nature
  Phys. {\bf 2}, 81 (2006).

\bibitem{book} {\it The Physics of Semiconductor Microcavities}, edited by B. Deveaud (Wiley-VCH, Weinheim, 2007).


\bibitem{rmp} H. Deng, H. Haug, and Y. Yamamoto, {\it Exciton-polariton
  Bose-Einstein condensation}, Rev. Mod. Phys. {\bf 82}, 1489 (2010).

\bibitem{rmp2}  I. Carusotto and C. Ciuti, {\it Quantum
  fluids of light},   Rev. Mod. Phys. {\bf 85}, 299 (2013).



\bibitem{deng-science} H. Deng, G. Weihs, C. Santori, J. Bloch,
  Y. Yamamoto, {\it Condensation of Semiconductor
Microcavity Exciton Polaritons}, Science {\bf 298}, 199 (2002).


\bibitem{bec1} H. Deng, G. Weihs, D. Snoke, J. Bloch, and Y. Yamamoto,
  {\it Polariton lasing vs. photon lasing in a semiconductor
  microcavity}, Proc. Natl. Acad. Sci. U.S.A. {\bf 100}, 15318 (2003).


\bibitem{bec3} J. Kasprzak, M. Richard, S. Kundermann, A. Baas,
  P. Jeambrun, J. M. J. Keeling, F. M. Marchetti, M. H. Szymaska,
  R. Andr\'e, J. L. Staehli, V. Savona, P. B. Littlewood, B. Deveaud,
  and L. S. Dang, {\it Bose-Einstein condensation of exciton
  polaritons}, Nature {\bf 443}, 409 (2006). 


\bibitem{bec4} R. Balili, V. Hartwell, D. Snoke, L. Pfeiffer,
  K. West, {\it Bose-Einstein Condensation of Microcavity Polaritons in a
  Trap}, Science {\bf 316}, 1007 (2007).




\bibitem{superfluid} A. Amo, J. Lefr\`ere, S. Pigeon, C. Adrados,
  C. Ciuti, I. Carusotto, R. Houdr\'e, E. Giacobino, and A. Bramati,
  {\it Superfluidity of polaritons in semiconductor microcavities},
  Nature Phys. {\bf 5}, 805 (2009).


\bibitem{vort1}   
K. G. Lagoudakis, M. Wouters, M. Richard, A. Baas,
  I. Carusotto, R. Andr\'e, L. S. Dang, and B. Deveaud-Pl\'edran, 
{\it Quantized vortices in an exciton–polariton condensate}, Nature
Phys. {\bf 4}, 706  (2008) ; D. Sanvitto,	 F. M. Marchetti,
  M. H. Szyma\'nska,	 G. Tosi,	 M. Baudisch,	 F. P. Laussy,
  D. N. Krizhanovskii,	 M. S. Skolnick,	 L. Marrucci,
  A. Lema\'itre,	 J. Bloch,	 C. Tejedor, and L. Vi\~na, 
  {\it Persistent currents and quantized vortices in a polariton superfluid},
Nature Phys. {\bf 6}, 527 (2010); K. G. Lagoudakis,
T. Ostatnick\'y, A. V. Kavokin, Y. G. Rubo, R. Andr\'e, and
B. Deveaud-Pl\'edran, {\it Observation of Half-Quantum Vortices in an
Exciton-Polariton Condensate}, Science {\bf 326}, 974 (2009). 




\bibitem{joseph} C. W. Lai, N. Y. Kim, S. Utsunomiya, G. Roumpos,
  H. Deng, M. D. Fraser, T. Byrnes, P. Recher, N. Kumada, T. Fujisawa,
  and Y. Yamamoto, {\it Coherent zero-state and $\pi$-state in an
  exciton-polariton condensate array}, Nature {\bf 450}, 529  (2007).

\bibitem{soliton} A. Amo, S. Pigeon, D. Sanvitto, V. G. Sala,
  R. Hivet, I. Carusotto, F. Pisanello, G. Lem\'enager, R. Houdr\'e,
  E. Giacobino, C. Ciuti, and A. Bramati, {\it Polariton Superfluids
  Reveal Quantum Hydrodynamic Solitons}, Science {\bf 332}, 1167-1170
  (2011).


\bibitem{butov} L. V. Butov and A. V. Kavokin, {\it The behaviour of exciton-polaritons}, Nature Photon. {\bf
    6}, 2 (2012); B. Deveaud-Pl\'edran, Nature Photon. {\bf
    6}, 205 (2012).

\bibitem{snoke3} B. Nelsen, G. Liu, M. Steger, D. W. Snoke, R. Balili,
  K. West, and L. Pfeiffer, {\it Dissipationless Flow and Sharp
    Threshold of a Polariton Condensate with Long Lifetime},
  Phys. Rev. X {\bf 3}, 041015 (2013); M. Steger, G. Liu, B. Nelsen,
  C. Gautham, D. W. Snoke, R. Balili, L. Pfeiffer, and K. West, 
  {\it Long-range ballistic motion and coherent flow of long-lifetime
    polaritons}, Phys. Rev. B {\bf 88}, 235314 (2013).


\bibitem{atac} A. Imamo$\bar{{\rm g}}$lu,
  R. J. Ram, S. Pau, and Y. Yamamoto, {\it Nonequilibrium condensates and
  lasers without inversion: Exciton-polariton lasers}, Phys. Rev. A {\bf 53}, 4250 (1996). 


\bibitem{electric} C. Schneider, A. Rahimi-Iman, N. Y. Kim,
  J. Fischer, I. G. Savenko, M. Amthor, M. Lermer, A. Wolf, L.
  Worschech, V. D. Kulakovskii, I. A. Shelykh,
  M. Kamp, S. Reitzenstein, A. Forchel, Y.
  Yamamoto, and S. H\"ofling, {\it An electrically pumped polariton
  laser}, Nature {\bf 497}, 348 (2013); P. Bhattacharya,
  B. Xiao, A. Das, S. Bhowmick, and J. Heo, {\it Solid State Electrically
  Injected Exciton-Polariton Laser}, Phys. Rev. Lett. {\bf
    110}, 206403 (2013).

\bibitem{yang3} S. J. Yang and S. John, {\it Coherence and antibunching in a trapped interacting Bose-Einstein condensate},
  Phys. Rev. B {\bf 84}, 024515 (2011).



\bibitem{transistor} D. Ballarini, M. De Giorgi, E. Cancellieri,
  R. Houdr\'e, E. Giacobino, R. Cingolani, A. Bramati, G. Gigli, and
  D. Sanvitto, {\it All-optical polariton transistor}, Nat.
  Commun. {\bf 4}, 1778 (2013).

\bibitem{diode} H. S. Nguyen, D. Vishnevsky, C. Sturm, D. Tanese,
  D. Solnyshkov, E. Galopin, A. Lema\^{i}tre, I. Sagnes, A. Amo,
  G. Malpuech, and J. Bloch, {\it Realization of a Double-Barrier
  Resonant Tunneling Diode for Cavity Polaritons},
  Phys. Rev. Lett. {\bf 110}, 236601 (2013). 


\bibitem{switch-th} S. John and T. Quang, {\it Collective Switching and
  Inversion without Fluctuation of Two-Level Atoms in Confined
  Photonic Systems}, Phys. Rev. Lett. {\bf 78}, 1888-1891 (1997).



\bibitem{fast-switch} A. Amo, T. C. H. Liew, C. Adrados, R. Houdr\'e,
  E. Giacobino, A. V. Kavokin, and A. Bramati, {\it Exciton-polariton
    spin switches}, Nature Photon. {\bf 4}, 361-366 (2010); R. Cerna,
  Y. L\'eger, T. K. Para\"iso, M. Wouters, F. Morier-Genoud,
  M. T. Portella-Oberli, and B. Deveaud, {\it Ultrafast tristable spin
  memory of a coherent polariton gas}, Nat. Commun. {\bf
    4}, 2008 (2013).























\bibitem{rt} A. Kavokin, {\it Polaritons: the rise of the bosonic
    laser}, Nature Photon. {\bf 7}, 591-592 (2013); D. Snoke,
  {\it Microcavity polaritons: A new type of light switch}, Nature
  Nanotech. {\bf 8}, 393-395 (2013).


\bibitem{gan} S. Christopoulos, G. Baldassarri H. von
  H\"ogersthal, A. J. D. Grundy, P. G. Lagoudakis, A. V. Kavokin, 
  J. J. Baumberg, G. Christmann, R. Butt\'e, E. Feltin, J.-F. Carlin,
  and N. Grandjean {\it Room-Temperature Polariton Lasing in
  Semiconductor Microcavities},  Phys. Rev. Lett. {\bf 98}, 126405 (2007); 
  Gabriel Christmann, R. Butt\'e, Eric Feltin,
  J.-F. Carlin, and Nicolas Grandjean, {\it Room temperature polariton
  lasing in a GaN/AlGaN multiple quantum well microcavity},
  Appl. Phys. Lett. {\bf 93}, 051102 (2008). 

\bibitem{zno} Ying-Yu Lai, Yu-Pin Lan and
  Tien-Chang Lu, {\it Strong light-matter interaction in ZnO
  microcavities}, Light: Science \& Applications {\bf 2}, e76 (2013)

\bibitem{zno2}
  F. Li {\sl et al.}, {\it From Excitonic to Photonic Polariton
    Condensate in a ZnO-Based Microcavity}, Phys. Rev. Lett. {\bf 110}, 196406 (2013).

\bibitem{organic} D. G. Lidzey, D. D. C. Bradley,
  T. Virgili, A. Armitage, M. S. Skolnick, and S. Walker, {\it Room
    Temperature Polariton Emission from Strongly Coupled Organic
    Semiconductor Microcavities}, Phys. Rev. Lett. {\bf 82}, 3316 
  (1999); S. K\'ena-Cohen and S. R. Forrest, {\it Room-temperature
    polariton lasing in an organic single-crystal microcavity},
  Nature Photon. {\bf 4}, 371  (2010).


\bibitem{saturation} S. Schmitt-Rink, D. S. Chemla, and
  D. A. B. Miller, {\it Theory of transient excitonic optical
    nonlinearities in semiconductor quantum-well structures}, Phys. Rev. B {\bf 32}, 6601 (1985).


\bibitem{gan-ib} R. Butt\'e, G. Christmann, E. Feltin, J.-F. Carlin,
  M. Mosca, M. Ilegems, and N. Grandjean, {\it Room-temperature
    polariton luminescence from a bulk GaN microcavity}, Phys. Rev. B {\bf 73},
  033315 (2006).


\bibitem{zno-ib} R. Schmidt-Grund, B. Rheinl\"ander, C. Czekalla,
  G. Benndorf, H. Hochmuth, M. Lorenz, and M. Grundmann, {\it
    Exciton-polariton formation at room temperature in a planar ZnO
    resonator structure}, Appl. Phys. B
  {\bf 93}, 331 (2008).


\bibitem{gan-coupl} G. Christmann, R. Butt\'e, E. Feltin, A. Mouti,
  P. A. Stadelmann, A. Castiglia, J.-F. Carlin, and N. Grandjean, {\it
  Large vacuum Rabi splitting in a multiple quantum well GaN-based
  microcavity in the strong-coupling regime}, Phys. Rev. B {\bf 77}, 085310 (2008).


\bibitem{zno-coupl} J.-R. Chen, T.-C. Lu, Y.-C. Wu, S.-C. Lin,
  W.-R. Liu, W.-F. Hsieh, C.-C. Kuo, and C.-C. Lee, {\it Large vacuum
    Rabi splitting in ZnO-based hybrid microcavities observed at room temperature},
  Appl. Phys. Lett. {\bf 94}, 061103 (2009).


\bibitem{exp-dEx} G.V. Astakhov, V.A. Kosobukin, V.P. Kochereshko,
  D.R. Yakovlev, W. Ossau, G. Landwehr, T. Wojtowicz, G. Karczewski,
  and J. Kossut, {\it Inhomogeneous broadening of exciton lines in
    magneto-optical reflection from CdTe/CdMgTe quantum wells}, Eur. Phys. J. B {\bf 24}, 7-13 (2001).

\bibitem{weitz1} J. Klaers, J. Schmitt, F. Vewinger, and M. Weitz, {\it
    Bose-Einstein condensation of photons in an optical microcavity},
  Nature {\bf 468}, 545 (2010).


\bibitem{weitz2} J. Schmitt, T. Damm, D. Dung,
   F. Vewinger, J. Klaers, and M. Weitz, {\it Observation of
     Grand-Canonical Number Statistics in a Photon Bose-Einstein
     Condensate}, Phys. Rev. Lett. {\bf 112}, 030401 (2014); J. Klaers,
   J. Schmitt, T. Damm, F. Vewinger, and M. Weitz, {\it Bose-Einstein
     condensation of paraxial light}, Appl. Phys. B {\bf 105}, 17 (2011).

\bibitem{snoke-review} D. W. Snoke, {\it Polariton Condensation and
    Lasing}, in {\it Exciton Polaritons in Microcavities} (Springer
  Series in Solid State Sciences {\bf 172}), V. Timofeev and
  D. Sanvitto, eds. (Springer, 2012).






\bibitem{pbg1} S. John, {\it Strong localization of photons in certain
  disordered dielectric superlattices}, Phys. Rev. Lett. {\bf 58}, 2486 (1987).

\bibitem{pbg2} E. Yablonovitch, {\it Inhibited Spontaneous Emission in
  Solid-State Physics and Electronics}, Phys. Rev. Lett. {\bf 58}, 2059 (1987).

\bibitem{cavity-laser} O. Painter, R. K. Lee, A. Scherer, A. Yariv,
  J. D. O'Brien, P. D. Dapkus, and I. Kim, {\it Two-Dimensional Photonic
    Band-Gap Defect Mode Laser}, Science {\bf 284}, 1819 (1999).


\bibitem{noda} S. Ogawa, M. Imada, S. Yoshimoto, M.
  Okano, and S. Noda, {\it Control of Light Emission by 3D Photonic
    Crystals}, Science {\bf 305}, 227 (2004); S. Noda, M. Fujita,
  and T. Asano, {\it Spontaneous-emission control by photonic crystals
  and nanocavities}, Nature Photon. {\bf 1}, 449 (2007).


\bibitem{qd-cavity} K. Hennessy, A. Badolato, M. Winger, D. Gerace,
  M. Atat\"ure, S. Gulde, S. F\"alt, E. L. Hu, and A. Imamo$\bar{{\rm
      g}}$lu, {\it Quantum nature of a strongly coupled single quantum
  dot-cavity system}, Nature {\bf 445}, 896 (2007). 


\bibitem{Eb1} S. R. Jackson, J. E. Nicholls, W. E. Hagston,
  P. Harrison, T. Stirner, J. H. C. Hogg, B. Lunn, and
  D. E. Ashenford, {\it Magneto-optical study of excitonic binding
    energies, band offsets, and the role of interface potentials in
    CdTe/Cd$_{1-x}$Mn$_x$Te multiple quantum wells}, Phys. Rev. B {\bf 50}, 5392 (1994).


\bibitem{Eb2} R. Andr\'e, D. Heger, L. S. Dang, and Y. M. d'Aubign\'e,
  {\it Spectroscopy of polaritons in CdTe-based microcavities}, 
J. Cryst. Growth. {\bf 184/185}, 758-762 (1998). 

\bibitem{sat-exp} L. S. Dang, D. Heger, R. Andr\'e, F. B\oe{}uf,
  and R. Romestain, {\it Stimulation of Polariton Photoluminescence in
  Semiconductor Microcavity}, Phys. Rev. Lett. {\bf 81}, 3920
  (1998).



  







\bibitem{harrison} P. Harrison, {\it Quantum Wells, Wires and Dots}
  (Wiley, New York, 2001).

\bibitem{our} J.-H. Jiang and S. John, to be published.

\bibitem{Ab} S. Faure, T. Guillet, P. Lefebvre, T. Bretagnon, and
  B. Gil, {\it Comparison of strong coupling regimes in buk GaAs, GaN,
  and ZnO semiconductor microcavities}, Phys. Rev. B {\bf 78}, 235323 (2008).

\bibitem{mhhp} P. Harrison, F. Long, and W. E. Hagston, {\it Empirical
  pseudo-potential calculation of the in-plane effective masses of
  electron and holes of two-dimensional excitons in CdTe quantum wells}, Superlattices
  and Microstructures, {\bf 19}, 123 (1996).

\bibitem{dielectric}
D. T. F. Marple and H. Ehrenreich, {\it Dielectric Constant Behavior
  Near Band Edges in CdTe and Ge}, Phys. Rev. Lett. {\bf 8}, 87
(1962).

\bibitem{mgte-diel} {\it Handbook on Physical Properties of
    Semiconductors}, Vol. {\bf 3}, p.62, ed. S. Adachi (Kluwer
  Academic, Boston, 2004).

\bibitem{noda1} S. Noda, K. Tomoda, N. Yamamoto, and A. Chutinan, {\it
    Full Three-Dimensional Photonic Bandgap Crystals at Near-Infrared
    Wavelengths}, Science {\bf 289}, 604 (2000); S. Noda, M. Imada, M. Okano, S. Ogawa, M. Mochizuki,
   and A. Chutinan, {\it Semiconductor Three-Dimensional and
     Two-Dimensional Photonic Crystals and Devices}, IEEE J. Quantum Electron. {\bf 38}, 726 (2002).

\bibitem{noda2} S. Noda, N. Yamamoto, H. Kobayashi, M. Okano, and
  K. Tomoda, {\it Optical properties of three-dimensional photonic
    crystals based on III-V semiconductors at infrared to
    nrea-infrared wavelengths}, Appl. Phys. Lett. {\bf 75}, 905
  (1999).


\bibitem{high-q} Y. Takahashi, H. Hagino, Y. Tanaka, B.-S. Song,
  T. Asano, and S. Noda, {\it High-Q nanocavity with a 2-ns photon
  lifetime}, Opt. Express {\bf 15}, 17206 (2007).



\bibitem{pl-decay} M. O'Neill, M. Oestreich, W. W. R\"uhle, and
  D. E. Ashenford, {\it Exciton radiative decay and homogeneous
    broadening in CdTe/Cd$_{0.85}$Mn$_{0.15}$Te multiple quantum wells}, Phys. Rev. B {\bf 48}, 8980 (1993).

\bibitem{therm} R. Andr\'e, F. Boeuf, D. Heger, L. S. Dang,
  R. Romestain, J. Bleuse, and M. M\"uller, 
  {\it Cavity-polariton effects in II-VI microcavities}, Acta Physica Polonica A,
  {\bf 96}, 511 (1999).

\bibitem{infrared-phc} E. Nelson {\sl et al.}, {\it Epitaxial growth
    of three-dimensionally architecture optoelectronic devices}, Nature Mater. {\bf
    10}, 676 (2011).


\bibitem{kim} T. J. Kim, Y. D. Kim, and J. Kossut, {\it Study on the
    Dielectric Function of the CdMgTe Alloy by Using Vacuum
    Ultraviolet Spectroscopic Ellipsometry}, J. Korean
  Phys. Soc. {\bf 49}, 1156 (2006).

\bibitem{madelung} O. Madelung, {\it Semiconductors: Data Handbook}, 3rd ed. (Springer, Berlin, 2004).

\bibitem{group} M. S. Dresselhaus, G. Dresselhaus, and A. Jorio, {\it
    Applications of Group Theory to the Physics of Solids}, 1st ed. (Springer, 2008). 



\bibitem{mpb} We used the MIT photonic bands package to do the
  calculation. See http://ab-initio.mit.edu/wiki/index.php/MIT\_Photonic\_Bands


\bibitem{yang1} S. John and S. J. Yang, {\it Electromagnetically Induced
  Exciton Mobility in a Photonic Band Gap}, Phys. Rev. Lett. {\bf 99},
  046801 (2007).

\bibitem{yang2} S. J. Yang and S. John, {\it Exciton dressing and capture
  by a photonic band edge}, Phys. Rev. B {\bf 75}, 235332 (2007).

\bibitem{qw-broad} V. Savona, L. C. Andreani, P. Schwendimann,
  A. Quattropani, {\it Quantum Well Excitons in Semiconductor
    Microcavities: Unified Treatment of Weak and Strong Coupling
    Regimes}, Solid State Commun. {\bf 93}, 733 (1995).


\bibitem{lz1} B. Kuhn-Heinrich, W. Ossau, H. Heinke, F. Fischer,
  T. Litz, A. Waag, and G. Landwehr, {\it Optical investigation of
    confinement and strain effects in CdTe/(CdMg)Te quantum wells}, Appl. Phys. Lett. {\bf 63}, 2932
  (1993).

\bibitem{lz2} R. Spiegel, G. Bacher, K. Herz, A. Forchel, T. Litz,
  A. Waag, and G. Landwehr, {\it Recombination and thermal emission of
  excitons in shallow CdTe/Cd$_{1-x}$Mg$_x$Te quantum wells}, Phys. Rev. B {\bf 53}, 4544 (1996).

\bibitem{ovi} O. Toader, M. Berciu, and S. John, {\it Photonic Band Gaps
  Based on Tetragonal Lattices of Slanted Pores},
  Phys. Rev. Lett. {\bf 90}, 233901 (2003); O. Toader and S. John,
  {\it Slanted-pore photonic band-gap materials}, Phys. Rev. E {\bf 71}, 036605 (2005).


\bibitem{auger} M. A. Kinch, M. J. Brau, and A. Simmons, {\it
    Recombination mechanisms in 8-14~$\mu$ HgCdTe}, 
  J. Appl. Phys. {\bf 44}, 1649 (1973); J. Bajaj, S. H. Shin,
  J. G. Pasko, and M. Khoshnevisan, {\it Minority carrier lifetime in
    LPE Hg$_{1-x}$Cd$_x$Te}, J. Vac. Sci. Tech. A {\bf 1}, 1749 (1983).


\bibitem{srh} W. Shockley and W. T. Read, Jr., {\it Statistics of the
    Recombinations of Holes and Electrons}, Phys. Rev. {\bf 87},
  835 (1952); R. N. Hall, {\it Electron-Hole Recombination in Germanium}, Phys. Rev. {\bf 87}, 387 (1951);








\bibitem{multi2} T. K. Para\"iso, M. Wouters, Y. L\'eger,
  F. Morier-Genoud, and B. Deveaud-P\'edran, {\it Multistability of a
    coherent spin ensemble in a semiconductor microcavity}, Nature Mater. {\bf 9},
  655 (2010).



\bibitem{trapping} A. Das, P. Bhattacharya, J. Heo, A. Banerjee, and
  W. Guo, {\it Polariton Bose-Einstein condensate at room temperature
    in an Al(Ga)N nanowire-dielectric microcavity with a spatial
    potential trap}, Proc. Natl. Acad. Sci. USA {\bf 110}, 2735 (2013).


\bibitem{onsager} O. Penrose and L. Onsager, {\it Bose-Einstein
    Condensation and Liquid Helium}, Phys. Rev. {\bf 104}, 576
  (1956). 










\bibitem{ketterlepra} W. Ketterle and N. J. van Druten, {\it
    Bose-Einstein condensation of a finite number of particles trapped
  in one or three dimensions}, Phys. Rev. A
  {\bf 54}, 656 (1996).

\bibitem{wagner} N. D. Mermin and H. Wagner, {\it Absence of
    Ferromagnetism or Antiferromagnetism in One- or Two-Dimensional
    Isotropic Heisenberg Models}, Phys. Rev. Lett. {\bf
    17}, 1133 (1966).






\bibitem{cdmnte} L. Safonova, R. Brazis, and R. Narkowicz, {\it
    Complex Dielectric Constant of Cd$_{0.8}$Mn$_{0.2}$Te Crystals
    Near the Fundamental Absorption Edge}, Lith. J. Phys. {\bf 44},
  421 (2004).

\bibitem{mhhz} L. S. Dang, G. Neu, and R. Romestain, {\it Optical
    detection of cyclotron resonance of electron and holes in CdTe}, Solid State
  Commun. {\bf 44}, 1187 (1982).



\bibitem{band-gap} 
S. G. Choi, Y. D. Kim, S. D. Yoo, D. E. Aspnes, I. Miotkowski, and
A. K. Ramdas, {\it Ellipsometric studies Cd$_{1-x}$Mg$_x$Te $(0\le
  x\le 0.5)$ alloys}, Appl. Phys. Lett. {\bf 71}, 249 (1997).


\bibitem{mgte-Eg} J.-H. Yang, S. Chen, W.-J. Yin, X. G. Gong,
  A. Walsh, and S.-H. Wei, {\it Electronic structure and phase
    stability of MgTe, ZnTe, CdTe, and their alloys in the $B3$, $B4$, and
  $B8$ structures}, Phys. Rev. B {\bf 79}, 245202 (2009).

\bibitem{Ecv} A. A. Kiselev, E. L. Ivchenko, A. A. Sirenko, T. Ruf,
  M. Cardona, D. R. Yakovlev, W. Ossau, A. Waag, and G. Landwehr,
  {\it Electron and hole $g$ factor anisotropy in CdTe/CdMgTe quantum wells}, 
  J. Cryst. Growth. {\bf 184/185}, 831 (1998).
















\end{thebibliography}
\end{document}